\documentclass{article}
\usepackage{M-PSI}  %

\usepackage{lipsum} 
\usepackage{xcolor}
\usepackage{ifthen}

\usepackage[pdfencoding=auto, pdfusetitle]{hyperref}
\hypersetup{
  hidelinks,
  urlcolor=red,
  pdftitle={``I don't want to break it'': An Exploration of Perceived Fragility in Shape-Changing Interfaces},
  pdfauthor={Eva Mackamul, Tom Maillard, No\'e Marceaul, Yelli Coulibaly, Julien Pansiot, Laurence Boissieux, Dominique Vaufreydaz, Anne Roudaut, C\'eline Coutrix},
  pdfkeywords={shape-changing interfaces, tangible interfaces, user perception, interface fragility, breakability},
}

\setbool{twocolumns}{true}

\title{``I don't want to break it'': An Exploration of Perceived Fragility in Shape-Changing Interfaces}

\setsubtitle{Perceived Fragility in Shape-Changing Interfaces}

\newcommand\Affiliation[2]{%
\setbox0=\hbox{#1}%
\hspace{0.02cm}#1\hspace{-\wd0}#2
}

\newcommand{\AuthorEntry}[4]{%
    \mbox{
      \hspace{0.1cm}%
      \ifthenelse{\equal{#4}{}}{%
        #1\Affiliation{#2}\textsuperscript{#3}%
      }{%
        \href{#4}{#1\Affiliation{#2}\textsuperscript{#3,\ExternalLink}}%
      }%
      \hspace{0.1cm}
    }%
}

\newcommand{\Description}[1]{}
\newcommand{\finalAdd}[1]{#1}
\newcommand{\finalChanged}[1]{#1}
\newcommand{\finalDel}[1]{}
\usepackage{xspace}
\usepackage{macros}
\usepackage{subcaption}
\usepackage{tcolorbox}
\usepackage{eurosym}
\usepackage{float}
\usepackage{booktabs}
\usepackage[table]{xcolor}
\usepackage{makecell}

\author{
\AuthorEntry{Eva Mackamul}{,}{1}{https://evamackamul.com/}
\AuthorEntry{Tom Maillard}{,}{1}{}
\AuthorEntry{No\'e Marceaul}{,}{1}{}
\AuthorEntry{Yelli Coulibaly}{,}{1}{}
\AuthorEntry{Julien Pansiot}{,}{2}{}
\AuthorEntry{Laurence Boissieux}{,}{2}{}
\AuthorEntry{Dominique Vaufreydaz}{,}{1}{https://research.vaufreydaz.org/}
\AuthorEntry{Anne Roudaut}{,}{3}{}
\AuthorEntry{C\'eline Coutrix}{}{1}{https://iihm.imag.fr/coutrix/}\\
\vspace{0.5em}
\textnormal{\normalsize{
$^1$ Univ. Grenoble Alpes, CNRS, Grenoble INP\textsuperscript{*}, LIG, 38000 Grenoble, France\\[0.2em] %
$^2$ Univ. Grenoble Alpes, CNRS, Inria, Grenoble INP\textsuperscript{*}, LJK, 38000 Grenoble, France\\[-0.2em]
$^3$ University of Bristol
}}
}

\usepackage{overpic}

\renewcommand{\subtitleFigure}
{
    \vspace{0.5cm}

      \centering
      \begin{overpic}[%
          width=\textwidth]{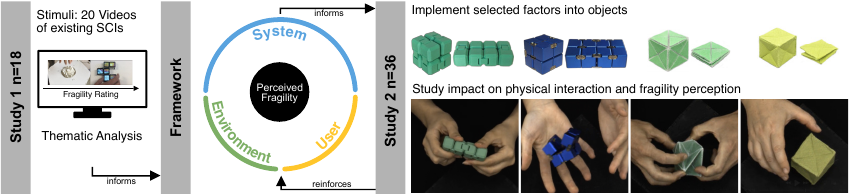}
          \put(0,0){\hyperref[section:study1]{\makebox(18.5,23)[lb]{}}}%
          \put(18.5,0){\hyperref[section:fragilityFramework]{\makebox(25,23)[lb]{}}}%
          \put(44,0){\hyperref[section:study2]{\makebox(56,23)[lb]{}}}%
      \end{overpic}
      \captionof{figure}{Study 1: Analysis of transcripts from 18 participants ranking videos of 20 existing SCIs according to their perceived fragility to examine fragility perceptions in general as well as which factors contribute to them. Framework: Resulting from the thematic analysis in study 1 we collated factors impacting the perception of fragility, sorting them into inherent \emph{system} factors, contextual and social \emph{environmental} factors and internal or interaction based \emph{user} factors. Study 2: Implementing selected factors found in the framework in shape-changing objects (showing 4 examples out of a total 12), we test their impact on interaction and fragility perceptions based on physical interaction.} 
      \label{fig:teaser}
      \vspace{0.5cm}
}

\begin{document}

\globalthanks{\textcopyright~2026 Copyright held by the owner/author(s). This is the author's version of the work. It is posted here for your personal use. Not for redistribution. The definitive version was published in CHI Conference on Human Factors in Computing Systems (CHI ’26), April 13 – 17, 2026, Barcelona, Spain, \href{https://doi.org/10.1145/3772318.3793422}{https://doi.org/10.1145/3772318.3793422}.}{}
\globalthanks{Institute of Engineering Univ. Grenoble Alpes}{*}

\begin{abstract}
Shape-Changing Interfaces (SCIs) dynamically alter their form, an inherent characteristic that introduces fragility into their design.
As a result, users’ perceptions of an interface’s fragility or its potential to move or break may influence their interaction, however the extent of this effect is unclear. 
To address this gap, we conducted a qualitative study (N = 18) using video stimuli showcasing 20 existing SCIs. 
Through thematic analysis, we identified key factors impacting perceived fragility and formalized these into a framework. 
We then conducted a second study (N = 36) for which we fabricated SCIs that varied across selected fragility-related dimensions. 
We recorded user interactions and compared how the selected dimensions shaped manipulation of the objects and how they were considered by users.
Together, these studies provide a structured foundational understanding of perceived fragility in SCIs and offer insights to enhance perceived robustness and inform future SCI development.
\end{abstract}
\keywords{shape-changing interfaces, tangible interfaces, user perception, interface fragility, breakability}

\section{Introduction}

Shape-changing interfaces (SCIs) are interactive systems capable of dynamically altering their physical form to provide new ways for users to interact with digital content and the physical environment \cite{alexander2018challenges, rasmussen2012shapeChangingReview}. By enabling tangible transformations such as 
bending \cite{youngwoo2015bendi}, 
rolling \cite{khalilbeigi2011xpaand}, 
expanding \cite{hirai2018xSlate, suzuki2019ShapeBots}, 
or twisting \cite{teyssier2019skinon, kim2019expandial}, 
SCIs open possibilities in areas including wearable technology, adaptive devices, robotics, and smart environments. 
They allow users to experience and manipulate interfaces through physical changes, creating richer and more engaging interaction experiences \cite{holman2008oui}.

Although SCIs promise versatility, their interactivity creates uncertainty that can amplify perceptions of fragility. Users may struggle to recognise available shape changes, hesitate to explore them, or misinterpret shape changes as unintended malfunctions (e.g. \cite{kwon2015fugaciousfilm}). Such unclear affordances may reduce confidence in the robustness of the system and reinforce concerns that experimenting with the interface might cause damage. 
These concerns are compounded by the materials and mechanisms that enable shape change. Hinges, flexible joints, soft materials, or inflatable elements are vulnerable to wear or breakage \cite{alexander2018challenges}, which can heighten user caution. 
\finalAdd{As \citeauthor{lee2023plantSci}~\cite{lee2023plantSci} note, users express \q{concern that their use of the artifacts would be damaging to the item itself, due to how they interpreted [its] form and materiality}, highlighting that fragility may \q{discourage usage}.}
Even widely adopted devices like smartphones frequently suffer breakage \cite{schaub2014broken}. What distinguishes SCIs, however, is that perceived fragility directly clashes with user expectations that novel interfaces should invite and withstand exploratory interaction.

Importantly, fragility manifests not only as a physical property but also as a \emph{user perception}, shaped by visual cues, anticipated material qualities, and expected mechanical behaviours. 
Existing studies outside the SCI domain demonstrate that perceived fragility strongly shapes user behaviour, confidence, and engagement. 
\finalChanged{For example, work on tangible objects demonstrates that physical dimensions affect fragility perception and user apprehension \cite{everitt2022investigating} while
studies involving blind users interacting with touchscreens highlight how unclear feedback can foster avoidance due to perceived fragility \cite{schaadhardt2021understanding}.}
These findings illustrate the impact of perceived fragility on interaction and highlight the importance of investigating how such perceptions influence user engagement in emerging interactive prototypes.
However, no research has directly examined how users perceive fragility in SCIs and how these perceptions affect their interactions. 
This lack of understanding limits our ability to design SCIs that are not only mechanically durable but also perceived as robust by users and subsequently invite exploration of available interactions.

To address the current lack of knowledge about perceived fragility in SCIs, we conducted a qualitative user study (N=18) where participants evaluated video stimuli showcasing 20 existing SCIs. 
Through thematic analysis of participant transcripts of a sorting task of these stimuli along a fragility spectrum, we identified 20 key factors influencing perceived fragility and formalised these into a framework. 
Building on these findings, we conducted a second study (N=36) in which we fabricated shape-changing objects that systematically varied across the selected fragility-related dimensions \emph{Material} and \emph{Demonstrated Handling} or \emph{Composition} and \emph{Movement}. 
Video recordings were analysed to examine how the selected dimensions affected interaction and perceived fragility. 
Results revealed that while some dimensions, like \emph{material} had an immediate effect on the perception of fragility, it was typically a combination of dimensions rather than isolated factors that informed participant reasoning.
Further, interaction analysis revealed that, while dimensions like \emph{movement} did not impact consciously articulated fragility perceptions, they did significantly alter behaviour patterns.

Together, these studies provide a structured foundational understanding of fragility in SCIs by revealing dimensions that inform the perception of fragility,
and exploring their impact on user behaviour.
In doing so, we offer a framework supporting the conscious consideration of fragility in the design of SCIs and subsequently encourage the development of interfaces that invite user engagement.

\section{Related Work}

\subsection{Shape-Changing Interfaces}
SCIs are computing systems that use the physical change of shape or materiality as input and/or output \cite{rasmussen2012shapeChangingReview}. 
As such, they are interactive systems that convey information, meaning, or emotion, and are controlled by computational processes, thus enabling them to respond autonomously or to user input \cite{alexander2018challenges}.
Their design dimensions, technical implementations, and possible usage contexts vary widely \cite{coelho2011shape}.
They are made utilising diverse materials including 
plastic \cite{visschedijk2022clipwidgets, dai2024morphmatrix, teyssier2018mobilimb}, 
rubber and silicon \cite{ teyssier2019skinon, sun2024magnedot, hirai2018xSlate}, 
paper \cite{kim2019expandial, guan2024waxpaper, suzuki2019ShapeBots}, 
metal \cite{nakagaki2016ChainFORM}, 
fabric \cite{haynes2024flextiles, sahoo2016tablehop}
or even water \cite{nakagaki2016hydromorph}
and animated through 
motors \cite{dai2024morphmatrix, zenner2019dragon, teyssier2018mobilimb}, 
magnets \cite{sun2024magnedot, tahouni2020NURBSforms, yasu2024magneSwift}, 
(air) pumps \cite{wang2024kipneu, chen2021pneuseries, pardomuan2024vabricbeads, morita2023inflatablemod}, 
environmental factors like heat \cite{haynes2024flextiles, zhong2023epomemory} or 
moisture \cite{raegrant2024excells, guan2024waxpaper} and 
physical user input \cite{teyssier2019skinon, kim2019expandial, sahoo2016tablehop}.
They also vary widely in scale, from adaptive environments \cite{teng2019tilepop, onishi2022waddlewalls, gronbaek2017proxemic} to fitting in the palm of ones hand \cite{sun2024magnedot, nicolae2024softbiomorph, liu2024breatHaptics}.
These different properties have been collated and examined with differing focal points in taxonomies including physical properties such as shape, orientation and material as well as movement speed and interaction method \cite{sturdee2018classification, rasmussen2012shapeChangingReview, kwak2014designspace, coelho2011shape, roudaut2013morphees}.

Notably, while popular approaches to shape-change exist, such as pin-based interfaces (\eg \cite{dai2024morphmatrix, nakagaki2016materiable, qian2024shapeit, lee2014stroke, nakagaki2017animastage, gao2024multimodalai, nakagaki2020transDock, nakagaki2019inforce, sturdee2023barChart}), which have been explored across various dimensions such as shape, material, movement and actuation, the majority of technologies utilised are novel. 
This novelty challenges users by requiring them to not only comprehend the system and its functionality but determine available and permissible interactions, oftentimes continuously as the physical shape of the interface morphs.

\subsection{Affordances of Shape-Changing Interfaces}
The challenge of interpreting both the information conveyed and interactions available in SCIs has been widely discussed in the context of affordances.
In particular, the concept of \emph{dynamic affordances} has been used to encapsulate the change in available or perceivable affordances in SCIs \cite{rasmussen2012shapeChangingReview, lopes2015affordancesPlus, follmer2015dynamic}.
They have been defined by \citeauthor{rasmussen2012shapeChangingReview}~\cite{rasmussen2012shapeChangingReview} as perceived interaction possibilities that change alongside the interfaces shape.
Meanwhile, \citeauthor{petersen2020affordances}~\cite{petersen2020affordances} argue that affordances may remain the same, while a change in state (shape) results primarily in a changed \emph{perception} of available functions and interactions.
Consequently, they argue that the main challenge arising from the dynamic nature of SCIs in regards to affordances, is to signify interaction possibilities as they become relevant.

However, the effectiveness of using aspects of visio-haptic representation of interfaces to signify interaction availability remains contentious.
While some studies found movement and shape changes to be especially suited to communicate affordances \cite{follmer2013inFORM}, others achieved mixed results \cite{economidou2021no} or found utilising system movement as an interaction signifier to be challenging \cite{tiab2016understanding}.
In particular \citeauthor{tiab2016understanding} found that  participants interpretations of subtle movement prior to interaction varied while larger movements deterred engagement and were perceived as a sign not to engage
\cite{tiab2016understanding}.
They concluded that affordance signifiers could not readily be mapped to physical shape change. These diverging results of interaction signifiers highlight that the variety of implementations of SCIs makes interaction, even if it is signified, challenging.
Further, the perception of shape-changes as a \emph{deterrent} shows that SCIs need to be designed in a way that encourages interaction as well as exploration of the interface where interaction possibilities are unclear.

\subsection{Interface Fragility}
There is a variety of terminology associated with how resistant interfaces are.
For example, 
``robustness'' is primarily used to describe how stable a system is internally \cite{shahrokni2013systematic, scarr2013robustness} while 
``durability'' is commonly used in the context of sustainability and long-term resistance and consistency \cite{odom2009understanding, remy2014addressing}. 

To our knowledge, interface fragility has not been explicitly studied in the context of SCIs or HCI, although the fragility of SCIs is a well-recognized issue, as highlighted in Alexander et al.’s grand challenge for SCIs \cite{alexander2018challenges}. We make the conscious decision to utilise the term ``fragility'' to emphasise our focus on \emph{subjectively perceived} susceptibility to damage due to accidents, misuse and human error. A rare example we found in this space is by \citeauthor{grafton2025breakability}~\cite{grafton2025breakability}, who investigated perceptions of “breakability”.
Their study examined how visual cues, such as the height of an object, influenced participants’ perception of drop height error in a task where they had to drop a sphere onto a piece of glass. However, this work was limited to screen-based simulations and did not address fragility in the context of SCIs.

\subsection{Factors Influencing Perception of Fragility}

While there are studies examining concepts adjacent to the perception of fragility, their focal points generally diverge. \citeauthor{gorniak2010manipulation}~\cite{gorniak2010manipulation} study how users may adapt to varying levels of fragility. They define a fragility index according to the rigidity of an object in relation to grip force needed to hold an object steady or collapse it rather than considering subjective perceptions. \citeauthor{kwon2015fugaciousfilm}'s~\cite{kwon2015fugaciousfilm} study of an ephemeral interface based on a soap film display highlighted how the fragile and transient nature of the system influenced user behaviour. Uncertainty about the stability or durability of the interface led to hesitation or strategic exploitation during interaction, emphasizing the need to understand perceived fragility in SCIs \cite{kwon2015fugaciousfilm}. 
However, while they examined how ephemeral materials impact interaction, they do not consider how users evaluate a given level of fragility.
Various studies have also considered how visual material properties impact users' motor behaviour while interacting \cite{ingvarsdottir2021material, strappini2024sustainable} without specifically focusing on their impact on perceived fragility.

Nonetheless, there are studies that consider factors that influence the perception of fragility.  
\citeauthor{grafton2025breakability} demonstrate that perceived fragility can be empirically examined through screen-based interactions, highlighting the importance of considering perceptual cues in the design of interactive devices, including SCIs \cite{grafton2025breakability}.
A study investigating interaction with tangible objects, such as rods, revealed that physical characteristics like height and width affect user perception and behaviour. Participants expressed apprehension toward taller and thinner targets, perceiving them as fragile and fearing breakage, which influenced their interaction \cite{everitt2022investigating}.
Similarly, research with blind users interacting with digital touchscreen devices showed that the absence of feedback created uncertainty about whether manipulations occurred. This uncertainty fostered a sense of fragility, making users apprehensive about interacting for fear of unintended changes \cite{schaadhardt2021understanding}. These findings highlight the critical role of perceived fragility in shaping user confidence and interaction behaviour.
Finally, studies on the visual perception of materials contrast physical, mechanical properties (shape, motion) and surface optical appearance (gloss, translucency, texture) and their impact on perceived qualities of materials, including reactions to force \cite{schmid2018shatter, ingvarsdottir2020visual}. 
However due to their focus on material perception, they do not consider fragility within an interactive context.

\section{Study 1 -- Determining Factors of Perceived Fragility}
\label{section:study1}

To examine factors shaping the perception of fragility in SCIs, we conducted a user study where participants viewed videos of existing SCIs and positioned them on a fragility axis while explaining their reasoning. 

\subsection{Interface Selection}
\label{section:videoSelection}
\begin{figure*}[t]
    \centering
    \includegraphics[width=\textwidth]{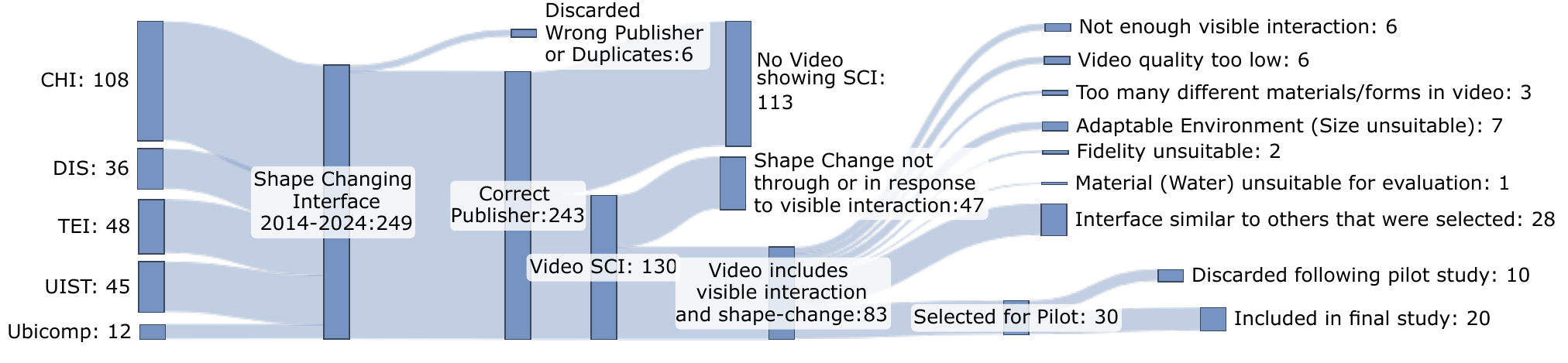}
    \vspace{-2em}
    \caption{Sankey plot showing initial results for ``shape-changing interface'' across conferences from 2014 to 2024 and how they were processed for video selection. 
    }
    \label{figure:videoSelection}
\end{figure*}
We used video stimuli, which enabled participants to observe a wide range of transformations while ensuring consistent presentation across all participants without the need for physical access to the devices. 
\finalChanged{In addition to providing a more thorough view of different SCIs, we elect this method to establish visual cues of fragility, since interfaces being perceived as fragile may discourage engagement prior to any physical interaction occurring.
Our choice is supported by related work showing that perceived breakability can be examined through screen-based presentations \cite{grafton2025breakability}. 
}
To capture a representative sample of existing SCIs, we curated video clips from prior publications (see \autoref{figure:videoSelection}). We focused on papers published between 2014 and 2024 in major ACM conferences, including CHI, DIS, TEI, UIST, and Ubicomp, that contained the term SCIs (243 papers in total) %
\begin{figure*}[h]
    \centering
      \begin{overpic}[%
      width=\textwidth]{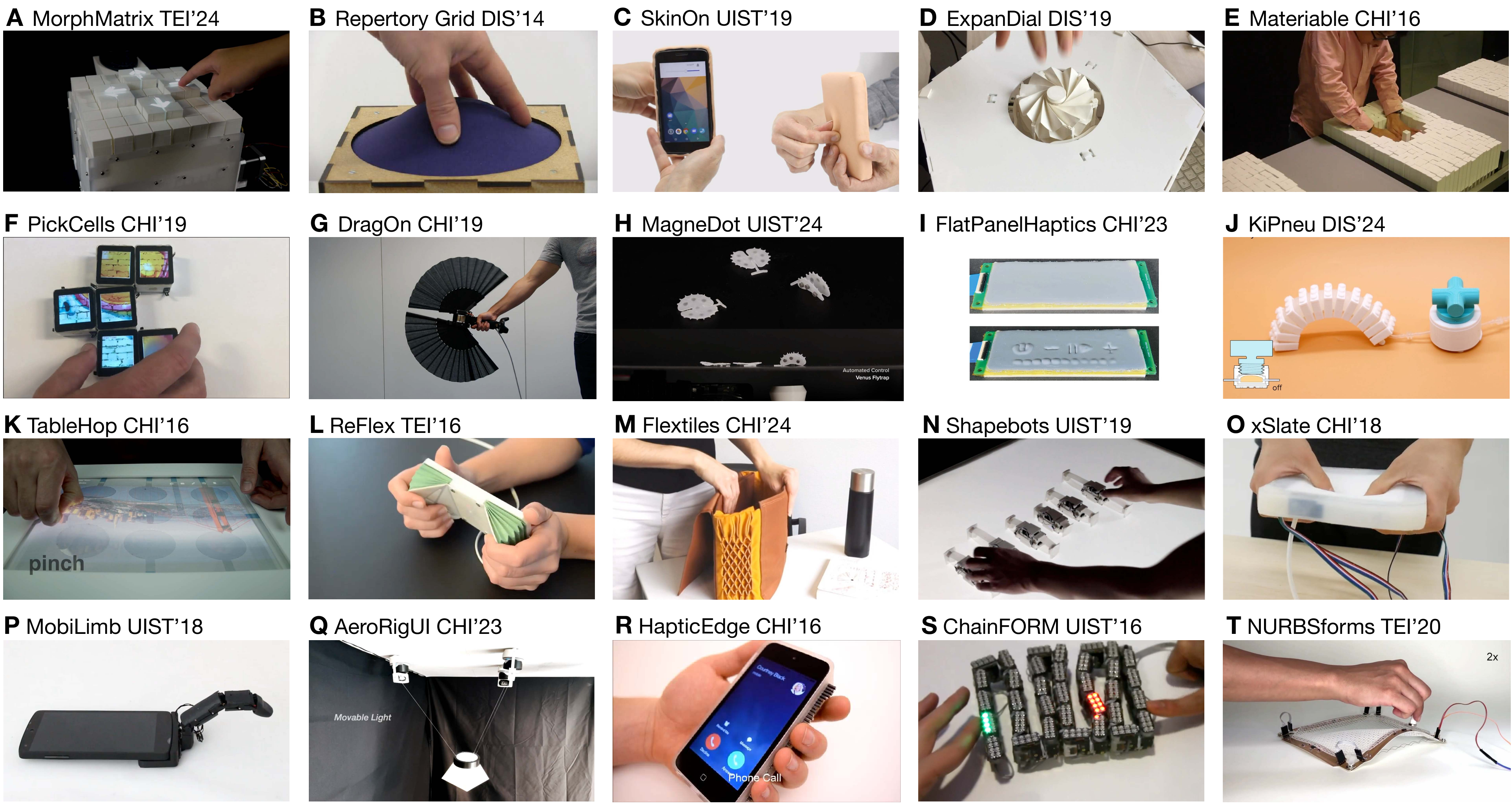}
      \put(0,41){\hyperlink{cite.dai2024morphmatrix}{\makebox(19,13)[lb]{}}}%
      \put(20.5,41){\hyperlink{cite.kwak2014designspace}{\makebox(19,13)[lb]{}}}%
      \put(40.5,41){\hyperlink{cite.teyssier2019skinon}{\makebox(19,13)[lb]{}}}%
      \put(60.8,41){\hyperlink{cite.kim2019expandial}{\makebox(19,13)[lb]{}}}%
      \put(81,41){\hyperlink{cite.nakagaki2016materiable}{\makebox(19,13)[lb]{}}}%
      \put(0,27.5){\hyperlink{cite.goguey2019pickcells}{\makebox(19,13)[lb]{}}}%
      \put(20.5,27.5){\hyperlink{cite.zenner2019dragon}{\makebox(19,13)[lb]{}}}%
      \put(40.5,27.5){\hyperlink{cite.sun2024magnedot}{\makebox(19,13)[lb]{}}}%
      \put(60.8,27.5){\hyperlink{cite.shultz2023flatpanelhaptics}{\makebox(19,13)[lb]{}}}%
      \put(81,27.5){\hyperlink{cite.wang2024kipneu}{\makebox(19,13)[lb]{}}}%
      \put(0,13){\hyperlink{cite.sahoo2016tablehop}{\makebox(19,13)[lb]{}}}%
      \put(20.5,13){\hyperlink{cite.rasmussen2016balancing}{\makebox(19,13)[lb]{}}}%
      \put(40.5,13){\hyperlink{cite.haynes2024flextiles}{\makebox(19,13)[lb]{}}}%
      \put(60.8,13){\hyperlink{cite.suzuki2019ShapeBots}{\makebox(19,13)[lb]{}}}%
      \put(81,13){\hyperlink{cite.hirai2018xSlate}{\makebox(19,13)[lb]{}}}%
      \put(0,0){\hyperlink{cite.teyssier2018mobilimb}{\makebox(19,13)[lb]{}}}%
      \put(20.5,0){\hyperlink{cite.yu2023aeroRigUI}{\makebox(19,13)[lb]{}}}%
      \put(40.5,0){\hyperlink{cite.jang2016hapticedge}{\makebox(19,13)[lb]{}}}%
      \put(60.8,0){\hyperlink{cite.nakagaki2016ChainFORM}{\makebox(19,13)[lb]{}}}%
      \put(81,0){\hyperlink{cite.tahouni2020NURBSforms}{\makebox(19,13)[lb]{}}}%
    \end{overpic}
    \vspace{-1em}
    \caption{SCIs selected as video stimuli alongside their assigned letter identifier. Stills are taken from the videos utilised and sourced from A \cite{dai2024morphmatrix},
B \cite{kwak2014designspace},
C \cite{teyssier2019skinon},
D \cite{kim2019expandial},
E \cite{nakagaki2016materiable},
F \cite{goguey2019pickcells},
G \cite{zenner2019dragon},
H \cite{sun2024magnedot},
I \cite{shultz2023flatpanelhaptics},
J \cite{wang2024kipneu},
K \cite{sahoo2016tablehop},
L \cite{rasmussen2016balancing},
M \cite{haynes2024flextiles},
N \cite{suzuki2019ShapeBots},
O \cite{hirai2018xSlate},
P \cite{teyssier2018mobilimb},
Q \cite{yu2023aeroRigUI},
R \cite{jang2016hapticedge},
S \cite{nakagaki2016ChainFORM},
T \cite{tahouni2020NURBSforms}.}
    \vspace{-1em}
    \Description{Summary overview of videos selected for study 1. Showing stills from the videos as well as their assigned letter, author and year and venue of publication.}
    \label{figure:allinterface}
\end{figure*}
and selected papers that included video presentations depicting interactions between a person and the shape changing interface (83 papers). 
The videos were assessed for video quality, interface size, visibility and material as well as interaction visibility and the similarity of the interfaces between the videos.
Given those considerations we selected 30 videos. Following pilot testing we selected 20 videos for the study.
Each video was assigned a letter identifier to facilitate discussion and analysis. The selected videos are as shown in \autoref{figure:allinterface}:
A \cite{dai2024morphmatrix},
B \cite{kwak2014designspace},
C \cite{teyssier2019skinon},
D \cite{kim2019expandial},
E \cite{nakagaki2016materiable},
F \cite{goguey2019pickcells},
G \cite{zenner2019dragon},
H \cite{sun2024magnedot},
I \cite{shultz2023flatpanelhaptics},
J \cite{wang2024kipneu},
K \cite{sahoo2016tablehop},
L \cite{rasmussen2016balancing},
M \cite{haynes2024flextiles},
N \cite{suzuki2019ShapeBots},
O \cite{hirai2018xSlate},
P \cite{teyssier2018mobilimb},
Q \cite{yu2023aeroRigUI},
R \cite{jang2016hapticedge},
S \cite{nakagaki2016ChainFORM},
T \cite{tahouni2020NURBSforms}.
For easy identification, the corresponding papers are marked with a {\textcolor{cyan}{$\clubsuit$}} symbol in the references. We extracted the interface and interaction segments from each video and removed audio to avoid distractions from voice-overs or background music.

\subsection{Experimental Design and Procedure}
\label{section:study1Procedure}
The experiment was conducted remotely on participants’ personal laptops or desktops. After giving informed consent, they completed a demographic survey (age, origin, occupation) and four 7-point Likert questions on technology experience (\autoref{figure:demographicLikert}) before opening a shared whiteboard link (Canva WebApp 1.103.0) and sharing their screen. The whiteboard 
displayed an arrow pointing to the right, with a label on the left (``Not at all Fragile'') and right (``Extremely Fragile'') ends. 
To its right two boxes were displayed, one with the ``Videos to be sorted'' and one below with the ``Annotation Materials'' including arrows, text boxes and post-it notes for easy access.
The ``Videos'' box displayed the videos included in the study in a randomised order for each participant. Participants familiarised themselves with Canva’s controls before watching the videos and sorting them along the axis while thinking aloud. They could add any number of annotations and were reminded there were no right or wrong placements. 
\finalAdd{To avoid biasing participants towards specific factors, no definition of fragility was provided. However, if participants asked for clarification, it was described as the consideration of how easily the interface may break.}
Sessions lasted about 90 minutes, with recording beginning from the start of screen sharing.

\begin{figure*}[t]
    \centering
    \begin{subfigure}[b]{0.52\textwidth}
        \includegraphics[width=\textwidth]{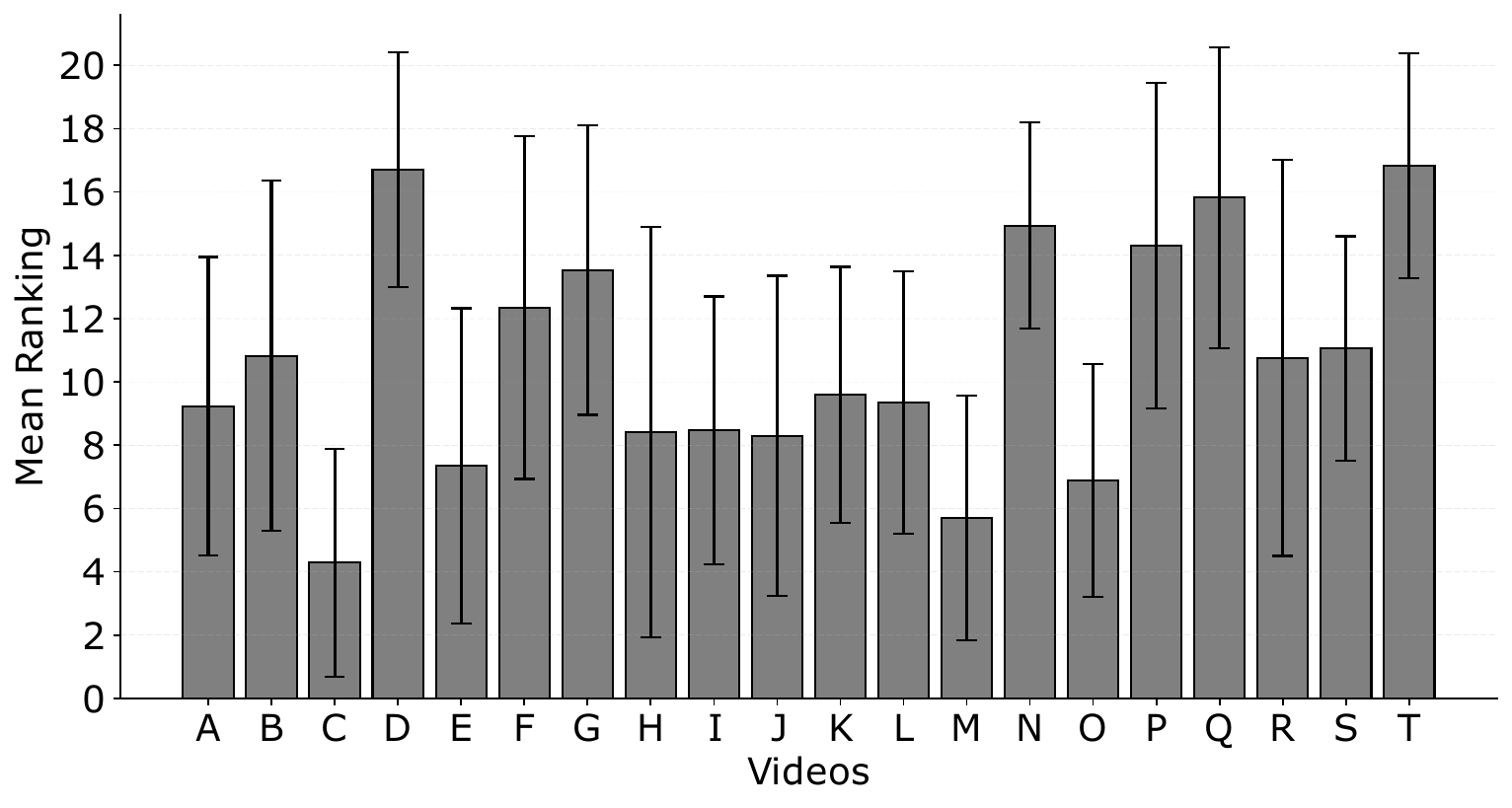}
        \caption{Mean fragility rankings and standard deviations}
        \label{subfig:videoRankingsMeanSD}
    \end{subfigure}%
    \begin{subfigure}[b]{0.48\textwidth}
        \includegraphics[width=\textwidth]{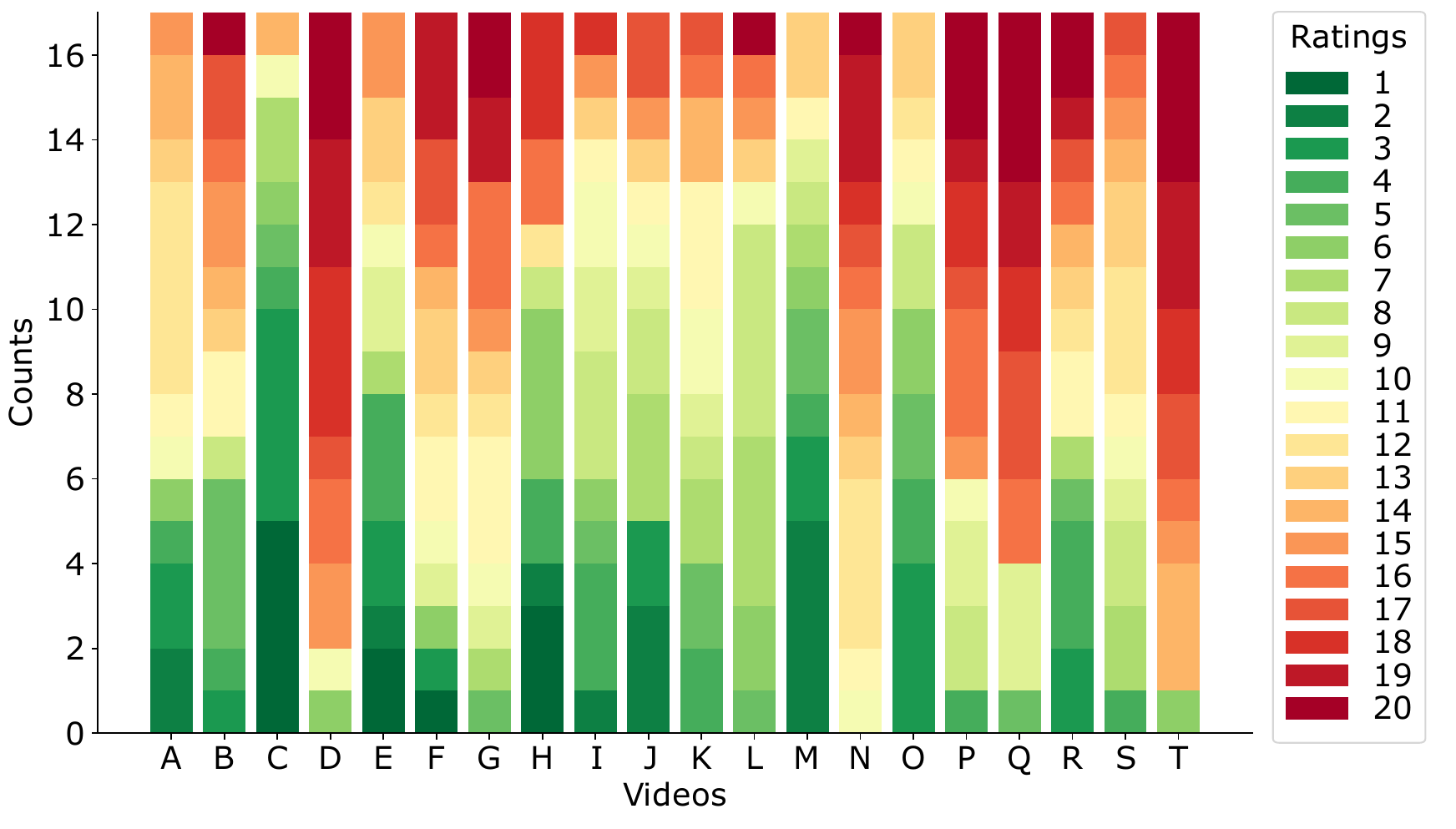}
        \caption{Absolute fragility rankings}
        \label{subfig:videoRankingsStackedVisual}
    \end{subfigure}
    \vspace{-1em}
    \caption{Graphs showing the fragility rankings of the different presented systems.}
    \Description{Graphs showing the fragility rankings of the different presented videos. Figure a on the left shows mean fragility rankings and standard deviations while figure b on the right shows the absolute fragility rankings as stacked bars.}
    \label{figure:videoRankings}
\end{figure*}

\subsection{Participants}
\begin{figure}[h]
    \centering
    \includegraphics[width=\linewidth]{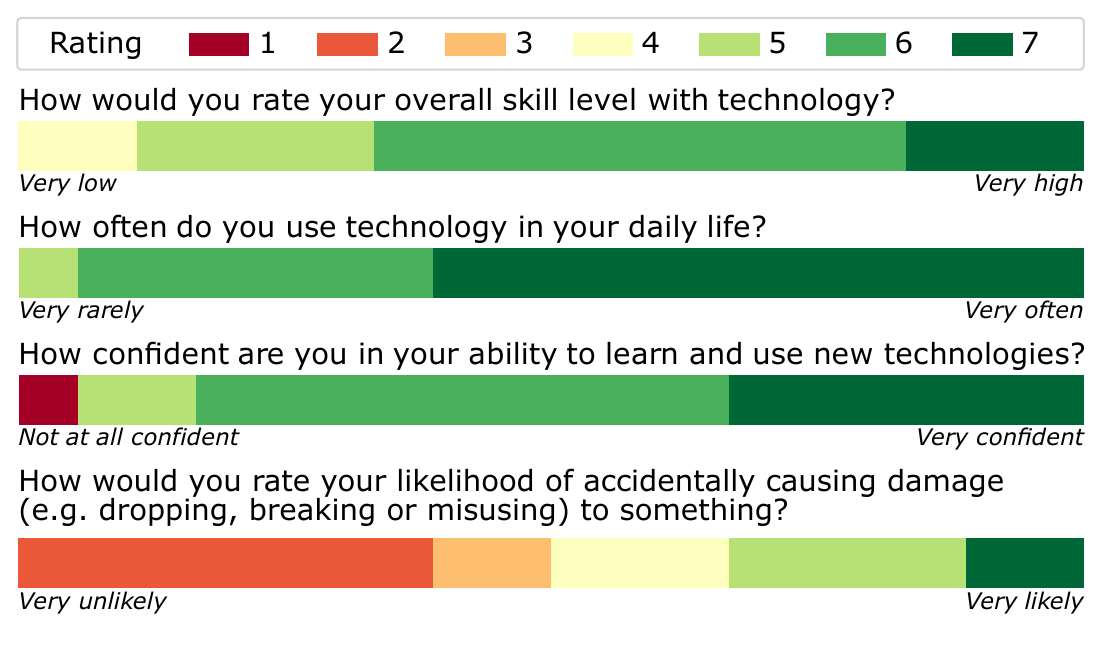}
    \caption{Visualisation of likert responses to technology experience questions in study 1 in stacked barplot. }
    \Description{Figure showing the Likert responses about technology experience from users}
    \label{figure:demographicLikert}
\end{figure}

We recruited 18 participants via word of mouth, with eligibility limited to adults fluent in English.
Age was reported in predefined ranges: 18–24 years (2), 25–34 years (14), 35–44 years (2), 45–54 years (0), and 55+ years (0).
10 participants self identified as women, 7 as men and 1 as non-binary. 
Occupations varied from Student (3), Education (4) and Technology or IT (6) to Design, Sports, Media, Research and Unemployed.
As \autoref{figure:demographicLikert} illustrates, 
participants self reported their skill level with technology (\statinfo{M=5.72, SD=0.89}), technology usage frequency (\statinfo{M=6.55, SD=0.62}), confidence in learning new technology (\statinfo{M=5.94, SD=1.39}) and likelihood to accidentally cause damage to something (\statinfo{M=3.67, SD=1.71}).
Participation was voluntary, and no compensation was provided.

\section{Study 1 - Results}

\subsection{\finalAdd{Fragility Definitions}}
\finalAdd{
While participants were not explicitly instructed to articulate their definition of fragility, some did.
Most commonly, they considered fragility in terms of their own ability to break an interface, e.g. \q{I don't think I could damage that} (P02).
In the same vein, some participants considered more concrete scenarios like P03 noting \q{it looks like as well if you sat on it, it would be in trouble.
That's kind of my scale here. Can you sit on it? Yes or no?}.
Notably P05 reiterated their definition multiple times, stating
\q{Maybe I need to rethink my definition of fragile. I think of it as brittle. Will it break?}, 
\q{Maybe I can simplify my definition of fragile [...] Like if I bend it, will it break?} 
and finally highlighting
\q{I just keep changing definitions because I don't think I have a very good definition of fragility}.
However, despite their struggle to define fragility, their definitions mirrored those of other participants focused around the potential breakability of an interface.
}

\subsection{Quantitative data}
\paragraph{Video ranking}
To determine the level of agreement between participants, we assigned videos a score according to their placement along the fragility axis starting with 20 for the most fragile interfaces.
When interfaces were considered equally breakable, we assigned the same score with the following score dropping according to the amount of videos (e.g. 20, 19, 19, 17, ...).
P09 placed the videos in only 3 distinctive groups and was subsequently removed for statistical analysis as an outlier. Their responses were nonetheless considered in the qualitative analysis.

Kendall's coefficient of concordance (\statinfo{W=0.36}) indicates moderate agreement.
This can be explained by the variety of systems presented as well as differing focal points and definitions regarding fragility.
The mean standard deviation across videos was \statinfo{SD=4.53} (see \autoref{figure:videoRankings}) with 6 systems' fragility rankings scoring a standard deviation higher than 5 (H (\statinfo{M=8.41, SD=6.48}), R (\statinfo{M=10.76, SD=6.26}), B (\statinfo{M=10.82, SD=5.53}), F (\statinfo{M=12.35, SD=5.42}), P (\statinfo{M=14.94, SD=5.14}), J (\statinfo{M=8.29, SD=5.06})).

There was no correlation of participants self-reported technology experience (\autoref{figure:demographicLikert}) and their rankings, neither was there a correlation between video presentation order and fragility rankings.

\begin{figure*}[ht]
    \centering
      \begin{overpic}[%
      width=0.55\textwidth]{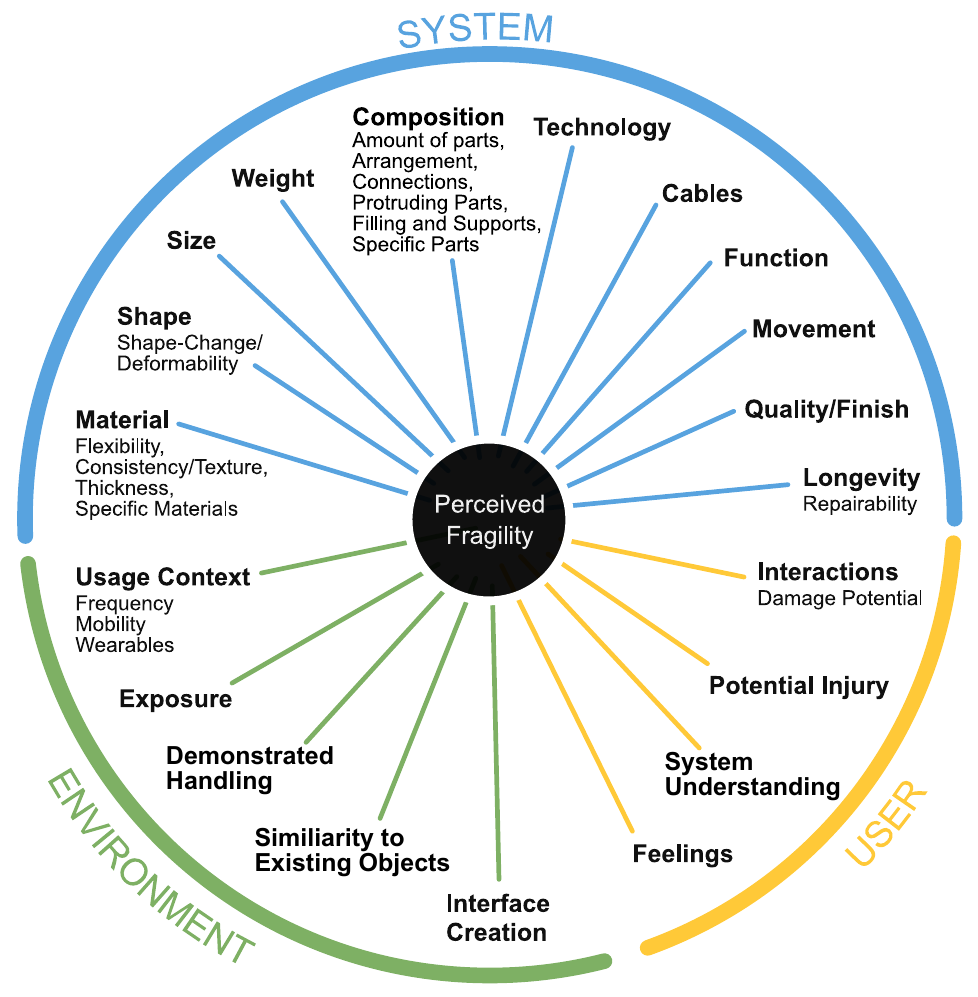}
      \put(8,48){\hyperref[section:thematicAnalysisMaterial]{\makebox(19,11)[lb]{}}}%
      \put(11.5,63){\hyperref[section:thematicAnalysisShape]{\makebox(15,7)[lb]{}}}%
      \put(16.5,75){\hyperref[section:thematicAnalysisSize]{\makebox(6,2)[lb]{}}}%
      \put(23,81){\hyperref[section:thematicAnalysisWeight]{\makebox(9,2.5)[lb]{}}}%
      \put(35,75){\hyperref[section:thematicAnalysisComposition]{\makebox(17,14.5)[lb]{}}}%
      \put(53.5,86){\hyperref[section:thematicAnalysisTechnology]{\makebox(14,2.5)[lb]{}}}%
      \put(66,79.5){\hyperref[section:thematicAnalysisCables]{\makebox(9,2.5)[lb]{}}}%
      \put(72.5,73){\hyperref[section:thematicAnalysisFunction]{\makebox(11,2.5)[lb]{}}}%
      \put(75.5,65.7){\hyperref[section:thematicAnalysisMovement]{\makebox(13,2.5)[lb]{}}}%
      \put(74.5,57.7){\hyperref[section:thematicAnalysisQuality]{\makebox(17,2.5)[lb]{}}}%
      \put(80,49){\hyperref[section:thematicAnalysisLongevity]{\makebox(13,4)[lb]{}}}%
      \put(8,35){\hyperref[section:thematicAnalysisUsageContext]{\makebox(17,8.5)[lb]{}}}%
      \put(11.5,28.7){\hyperref[section:thematicAnalysisExposure]{\makebox(12,2.5)[lb]{}}}%
      \put(16.5,20.7){\hyperref[section:thematicAnalysisHandling]{\makebox(17,4.5)[lb]{}}}%
      \put(25,12.5){\hyperref[section:thematicAnalysisSimilarity]{\makebox(20.5,4.5)[lb]{}}}%
      \put(44.8,5.7){\hyperref[section:thematicAnalysisCreation]{\makebox(11,4.5)[lb]{}}}%
      \put(63,13){\hyperref[section:thematicAnalysisFeelings]{\makebox(11,2.5)[lb]{}}}%
      \put(66,20){\hyperref[section:thematicAnalysisUnderstanding]{\makebox(18.5,4.5)[lb]{}}}%
      \put(71,30){\hyperref[section:thematicAnalysisInjury]{\makebox(18.5,2.5)[lb]{}}}%
      \put(76,39.5){\hyperref[section:thematicAnalysisInteractions]{\makebox(17,4.5)[lb]{}}}%
    \end{overpic}
    \caption{\finalChanged{Overview of themes identified in thematic analysis. System factors are inherent characteristics of the SCI linked to perceived fragility. Environmental factors depend on the surrounding context or social setting. User factors relate to users’ internal processes or interactions with the interface affecting fragility perception.}}
    \Description{Overview of themes identified in thematic analysis. System factors are inherent characteristics of the SCI linked to perceived fragility. Environmental factors depend on the surrounding context or social setting. User factors relate to users’ internal processes or interactions with the interface affecting fragility perception.}
    \label{figure:thematicAnalysisResults}
\end{figure*}

\paragraph{Clustering}
To see patterns in how participants ranked the videos, we clustered their rankings across videos using a K-means clustering approach.
This resulted in 2 clusters (Cluster 1: \statinfo{\finalChanged{P01, P02, P03, P04, P05, P07, P08,} P10, P11, P12, P13, P14, P16, P17, P18}; Cluster 2: \statinfo{\finalChanged{P06}, P15}).
Cluster 2 includes the only participants who diverged from a consensus, rating a video as fragile. \finalChanged{P06} diverged from common opinion by not considering interface D to be particularly fragile. This may be attributed to them considering the context in which the interface would be used, stating 
\q{I see this breaking at some point but not being too fragile. So it's like limited like controlled use. Because for example, comparing it to the phone, the use of this knob is going to be much more controlled, much more careful. I would say it's breaking much less.} 
\finalAdd{(P06)}.
This indicates, that they did not necessarily consider the interface to be less fragile than other participants but simply weighed the context in which interactions were likely to take place differently. P15 differed from the general rating of interface T by not considering it fragile. They reasoned that the suggested interaction lacked direct contact, saying  
\q{oh, yeah, you just point your magic wand magnet thing at one or the other of the little points} and judged the components less fragile 
\q{because you don’t touch them directly}. They further assumed shape changes were limited by design to avoid damage, explaining,  
\q{I assume that if there was like a limit to how much this can bend, for example, it would be kind of like programmed in the software. And so you could not like move it so far that it breaks}.
\finalChanged{
These examples highlight different perspectives when evaluating fragility. We therefore examine factors impacting those differences in a qualitative analysis.
}

\subsection{Thematic Analysis Overview}
\label{section:thematicAnalysis}
We collected 19h 43m of screen and audio recordings and transcribed sessions with MacWhisper 12.4 (Small English model). 
We then conducted a descriptive thematic analysis \cite{braun2006thematicanalysis} of participants’ think-aloud data to identify features associated with fragility (\autoref{figure:thematicAnalysisResults}). 
We refer to our approach as ``descriptive'' because our goal was to identify recurring factors that participants associated with fragility. This aligns with a more semantic-level analysis, where themes are closely tied to the surface content of the data. The coding was performed by two researchers. Three additional authors cross-validated the themes that were then discussed and refined collaboratively among the authors. 
\finalAdd{For a comprehensive overview of participant counts per theme please refer to \autoref{table:factorMentions}.}

\subsection{System Factors (Inherent Characteristics)}
\label{section:thematicAnalysisSystem}

\begin{tcolorbox}\textbf{Summary:}
    Participants highlighted material as a main factor affecting perceived fragility, considering flexibility (13 participants), consistency and texture (11), thickness (10), and specific material types (18). Beyond material, participants considered shape (17), including deformation and shape-change (18), size (15), weight (5), and composition 
    factors (18) such as number of parts, arrangement, connections and protruding parts, internal supports or filling and specific parts. Technology was considered by eight participants, with electronics, motors, sensors, and complex systems generally perceived as increasing fragility, whereas well-protected or mechanical components were seen as less fragile. Overall, material was foundational in shaping fragility perception, while other factors modulated this perception depending on context.
\end{tcolorbox}

\subsubsection{Material}
\label{section:thematicAnalysisMaterial}
Participants considered
the material of the interface.
When appraising the impact of material perception on fragility, they considered flexibility (\participants{P02-P05, P07, P08, P11-P15, P17, P18}), consistency and texture (\participants{P06-P09, P11, P12, P14-P18}), thickness (\participants{P01, P02, P09-P12, P14, P15-P17}) and opacity (\participants{P08, P16}) as well as specific material types (\participants{P01-P18}). Flexible and soft materials were often seen as durable, since they could bend or absorb shock without breaking. As one participant put it, an interface did not seem \q{fragile at all … the material seems like it’s flexible enough} (\participants{P11}). Yet others warned that flexibility/softness could signal delicacy, with concerns that such materials might deteriorate or \q{get caught on a lot of things} (\participants{P07}). 
Thickness was strongly associated with sturdiness, with thin materials viewed as fragile and thicker ones able to withstand force. Finally, participants drew on prior experience with familiar materials: paper (\participants{All -P06}) and glass (\participants{P02, P07, P11, P15}) were almost always described as fragile, metal (\participants{P02, P06, P07, P09-P12, P15, P17, P18}) and hard plastics (\participants{P01-P03, P07, P08, P10-P12, P15-P18}) as solid, while silicone (\participants{P01, P03, P08, P10, P12, P15-P18}) and textiles (\participants{P01-P04, P07-P10, P13-P18}) prompted more mixed views depending on their density or finish.

\subsubsection{Shape}
\label{section:thematicAnalysisShape}
A majority of participants (\participants{P01-P12, P14-P18}) considered the shape of the interface noting \q{the strength of a structure does depend on the shape} (\participants{P18}).
They described shapes that they did not consider fragile as bulky (\participants{P01, P09, P15}), compact (\participants{P04, P07}), girthy (\participants{P08}) and thick (\participants{P02, P11, P15}) as well as rectangular (\participants{P02, P05, P11, P18}), and shallow (\participants{P04}) or flat (\participants{P07, P10, P11}).
They described them as a \q{hunk of [material]} (\participants{P01}), \q{sturdy box} (\participants{P02}), a slab (\participants{P08, P11}), or \q{pillow shaped} (\participants{P16}), noting the shapes did not seem fragile because they had no openings (\participants{P05}), were compact and \q{less subject to tearing or breaking} (\participants{P07}), and had large flat surfaces (\participants{P07, P11}).
This is exemplified by P05 noting \q{It's like a rectangular block that's been cut up [...] how much more stable can something get?}. Inversely participants commented on shapes they considered fragile as thin (\participants{P02, P03, P06, P07, P11, P15, P16}) and long (\participants{P07, P15, P16}) noting that \q{there's more of a lever - there's probably more force involved} (\participants{P16}) when referring to long shapes. Similarly, they worried about height (\participants{P18}) and saw vertically expanding shapes as more fragile than horizontal ones (\participants{P08}). They also described an interface as fragile \q{because of the shape, because of the design, because they don't feel like a solid sheet} (\participants{P10}).

The majority of participants considered shape changes (\participants{P02-P04, P07, P13, P16, P18}) or the deformability of the interface (\participants{P01, P03-P05, P07, P12, P14-P17}) as a factor of fragility. Some saw resistance to twisting or bending as a sign of sturdiness, while others highlighted the opposite, describing shapes that \q{don’t feel that firm} (\participants{P14}) or noting that \q{the shape changing part feels a bit more fragile} (\participants{P07}). Deformable screens were often judged \q{more breakable than like a regular screen} (\participants{P18}), with certain states of a shape change seen as especially vulnerable. Concerns included permanent deformation and higher risk during transitions. Yet a few participants also considered deformability protective, reasoning that a screen able to bend \q{should take less damage} (\participants{P12}) if dropped. %

\subsubsection{Size}
\label{section:thematicAnalysisSize}
Most participants (\participants{P02-P04, P07, P08, P10-P18}) considered the size of the interface.
While the majority (\participants{P01, P03, P04, P07, P08, P12, P13, P17, P18}) considered smaller interfaces and parts as more fragile and larger interfaces as more sturdy, P02, P10 and P11 found both small and large sizes could reduce fragility and P14 and P18 argued that a small size could lead to less fragile interfaces.
Their arguments included that an interface may be \q{too small [...] too break it} (\participants{P02, P11, P14}), that smaller parts are \q{sturdier} (\participants{P18}) and big surfaces may have less support (\participants{P18}).
However, participants agreed that \q{the smaller it is the more vulnerable} (\participants{P11}) since \q{it's hard to make something that's so small and sturdy} (\participants{P15}).
They noted that at a larger size \q{it would be easier for me to perceive when I'm about to break something} (\participants{P15}) and that one wouldn't move larger interfaces around as much (\participants{P13}).

\subsubsection{Weight}
\label{section:thematicAnalysisWeight}
Participants noted the risk posed by weight in a hanging interface (\participants{P01, P02, P07}) and the increased tisk of dropping heavy interfaces (\participants{P14}), while lighter interfaces were seen as less fragile (\participants{P02}). However, P09 observed that extremely light objects, \q{like a leaf or paper} could also be fragile.

\subsubsection{Composition}
\label{section:thematicAnalysisComposition}
The composition of the system was universally considered to impact their fragility. This included the amount and arrangement of parts, their connections, protruding elements, and whether parts were hollow or supported. Participants reflected on these factors, including amount of parts (\participants{P01, P02, P04, P06–P08, P10–P15, P17, P18}), arrangement (\participants{P01, P02, P07, P08, P11, P13, P16–P18}), connections (\participants{P01–P03, P05–P10, P12–P18}), protruding parts (\participants{P01–P07, P09–P11, P13, P15–P18}), filling and supports (\participants{P01–P05, P07–P13, P15–P18}), and specific parts (\participants{P02–P18}).

Participants tended to equate more pieces and visible connections with higher fragility, whereas monolithic or well-supported structures were seen as sturdier. As one explained, \q{the number of stuff that are attached, not permanently to the surface, makes me feel that it’s very fragile} (\participants{P10}). Loose or protruding parts were often described as sources of anxiety, since they could break off or get caught during use: \q{they can get stuck in pretty much anything} (\participants{P07}). The nature of connections was a particular concern, with participants highlighting joints, magnets, hinges, or glue as both potential weak points and, when well designed, sources of strength. Hollow interiors or uncertain supports raised doubts, whereas protective casings and enclosures reassured participants. Finally, some linked fragility to specific elements such as wires, rods, or thin threads, while noting that modular systems could reduce perceived risk by localising damage.

\subsubsection{Technology}
\label{section:thematicAnalysisTechnology}
Participants considered the technological implementation (\participants{P01, P03, P08-P11, P15, P17}), in particular noting 
electronics (\participants{P1, P3, P11, P12, P15}), motors (\participants{P03, P07, P13, P17}), hardware (\participants{P11}) sensors (\participants{P09, P13, P17}) and \q{technology} (\participants{P08, P10}), as well as presence of electricity (\participants{P11}) as factors making an interface more fragile.
They stated they \q{[felt] like electronics have a higher likeliness of breaking or stopping to work} (\participants{P01}),  noting they are \q{more sensitive than other inventions} (\participants{P11}) and particularly highlighted small electronic parts (\participants{P11, P17}) and complex systems (\participants{P15}) as vulnerable. Participants noted \q{simple} electronics as less fragile (\participants{P01, P15}) and preferred systems in which the electronics were well hidden (\participants{P13, P15}).
Notably, P16 considered the \q{solid} electronic components of one interface as a factor lowering fragility.
Nonetheless, multiple participants (\participants{P01, P13, P15}) stated their preference for mechanical interfaces noting it is \q{always sturdier to have something that's mechanics based rather than electronic} (\participants{P01}) and \q{it's all based on this pneumatic thing and I feel like maybe I'm more comfortable with such non-electronic objects} (\participants{P15}).

\subsubsection{Cables}
\label{section:thematicAnalysisCables}
The presence of cables, wires, and tubes was considered almost universally (\participants{P01-P17}).
Participants were concerned about both, damaging the cables themselves (e.g. \q{these little cables could be torn apart} (\participants{P11}), \q{I'm afraid of breaking the cable.} (\participants{P03})) and damaging the system by disconnecting cables. E.g. \q{if the connection isn't working perfectly well, it just stops working} (\participants{P01}), \q{there are a lot of cables, and to me they seem easy to pull out or damage accidentally} (\participants{P02}). Participants noted the risk of tripping over cables or getting tangled.

\subsubsection{Function}
\label{section:thematicAnalysisFunction}
The intended function of an interface impacted it's perceived fragility for some participants.
While P11 considered an interface that could be used for multiple purposes likely to be more durable by design, P06 and P08 argued a multi-purpose interface was likely more prone to break.
Similarly, P08 considered interfaces with a clear function likely to be less fragile due to being optimised for their purpose and P17 noted that interfaces with \q{simple} functionality would not be fragile.
\finalDel{Finally, P03 noted that an interface who's function made it less likely to be touched often, would be less likely to break.}

\subsubsection{Movement}
\label{section:thematicAnalysisMovement}
Many participants (\participants{P01-P09, P11-P18}) considered the interface’s movement, focusing on independent movement before interaction or in response to user input. Movement was seen as both increasing (\participants{P01-P04, P07, P10, P13, P15, P16}) and decreasing (\participants{P01, P02, P04, P07, P14, P16}) fragility, depending on type. 
Movement decreasing fragility was described as \q{adaptable} (\participants{P14}), \q{reactive and autonomous} (\participants{P01, P09, P14}), \q{confident} (\participants{P16}), or having \q{elasticity and bounce} (\participants{P07}). Small movements (\participants{P05, P16, P18}), constrained movement (\participants{P15}), and system movements perceived as programmed rather than reactive (\participants{P09, P15}) were also seen as less fragile. P14 noted \q{organic} interfaces felt more fragile, stating \q{the minute they start looking very organic, they feel fragile to me}.

Conversely, some (\participants{P01, P02, P15, P16}) felt \q{less fragile, just because they don't have really moving parts} (\participants{P02}). Movement increasing fragility was described as \q{disjointed} (\participants{P06}), \q{erratic} and \q{unstable} (\participants{P12}), and \q{skittish} (\participants{P16}). P15 stated \q{even when it was moving [...] it seemed like it was struggling}, P16 noted \q{just didn't inspire confidence}, and P17 highlighted dangers of \q{imprecise} movements. P07 mentioned concerns about unreactive movements. Participants were also cautious interacting contrary to prior movements (\participants{P03, P11, P13, P16}), stating \q{it's something that's tried to resist your movement, I feel like it could easily break apart} (\participants{P03}) and \q{you shouldn't restrain the movement, other it risks breaking} (\participants{P16}). P03 added \q{The issue with that one is that you are touching the part that is moving}

\subsubsection{Quality/Finish}
\label{section:thematicAnalysisQuality}
The perceived finish of interfaces impacted the perception of fragility.
Participants considered systems as \q{well put together} (\participants{P04}), \q{well built} (\participants{P08}) and \q{quite sturdy based on its level of execution and its craftsmanship} (\participants{P01}), stating that \q{the way that it's manufactured doesn't offer a lot of weak points or possibility of breaking it} (\participants{P15}).
Notably, P01 expressed scepticism at a system looking sleek and \q{gimmicky [and] too good to be true}.
On the other side, several participants commented on systems that appeared very fragile due to their impressions as \q{prototypes} (\participants{P01, P02, P06, P07, P12-P17}) \finalChanged{referring to both appearance and movement (e.g. \q{the way that they're moving feels very "oh I am a prototype"} (\participants{P01})).}
P08 highlighted the monetary value of interfaces, stating \q{just [from] how much money it would take to build, [...] this is the least fragile} and \q{very fragile though. Also kind of easily replaceable, just a little bit cheap looking}.

\subsubsection{Longevity}
\label{section:thematicAnalysisLongevity}
Participants considered usage over time and the longevity of systems when evaluating their fragility (\participants{P01-P04, P06, P08, P12, P17}).
This largely took the form of noting systems would \q{break by themselves} eventually \q{just because you use them} (\participants{P17}).
They noted that even systems that are \q{kind of hard to break} were \q{likely to wear out} (\participants{P04}), especially for systems that are likely to be used a lot (\participants{P03, P06}).
In contrast P08 and P17 reasoned that some interfaces would be long-lasting despite frequent usage, making them especially robust. Participants not only considered the likelihood of the system breaking, but also how easy it would be to repair (\participants{P01, P02, P04, P06-P08, P12, P15-P17}). They highlighted that for some systems, breaking parts would not hinder usage (\participants{P02, P06, P07, P12, P15, P16}), while for others all parts are essential (\participants{P02, P06, P12}) as P06 describes, \q{this could be more resilient to losing certain aspects of it without breaking the whole device, where this one I feel like smaller issues could stop the whole process}.
Further, participants described interfaces as less fragile due to \finalAdd{being less likely to be touched often (\participants{P03}), }\q{easily replaceable parts} (\participants{P01, P04, P08, P15, P17}) or being easy to \q{put back together} or \q{repair} (\participants{P07, P08, P10, P12, P15}).

\subsection{Environmental Factors (Surrounding context or social setting)}
\label{section:thematicAnalysisEnv}

\begin{tcolorbox}\textbf{Summary:}
Participants’ perceptions of interface fragility were shaped by environmental and social factors. Usage context (16 participants), including location, mobility, and interaction frequency, influenced fragility judgments, with highly mobile or frequently used interfaces seen as more fragile. Exposure to elements such as water, dirt, temperature changes, and gravity increased perceived fragility (5), while resistance to these factors reduced it (3). Demonstrated handling in videos affected perceptions: rough handling suggested robustness, gentle handling increased fragility, and unclear cues caused doubt (16). Comparisons to familiar objects, most often touchscreens and smartphones, guided judgments (17), and the intentions or creation process behind interfaces also mattered, with well-designed or tested interfaces perceived as less fragile (6).
\end{tcolorbox}

\subsubsection{Usage Context}
\label{section:thematicAnalysisUsageContext}
Participants widely considered the context in which the interface would be used (\participants{P01–P09, P11–P13, P15–P18}), with some judging it more fragile depending on use, while others reasoned that context could reduce fragility. As one put it, \q{maybe the context counts too much for me here, but I think it’s more fragile because of that} (\participants{P12}), whereas another explained, \q{if you keep it in that setting, then I don’t think it’s fragile} (\participants{P13}). Participants linked fragility to frequency of use (\participants{P02, P03, P06, P08, P13, P16}), mobility (\participants{P02, P03, P06, P07, P11–P13, P15, P17}), and wearable contexts (\participants{P04, P06, P08, P11, P13, P15}). Frequent handling, especially of devices resembling phones, was seen as risky because such objects are \q{not something you are used to holding carefully} (\participants{P08}). Mobile systems were considered fragile due to the potential for dropping or snagging, while fixed installations were thought sturdier since they could not be easily displaced. %
Wearables elicited mixed views: some feared they would deteriorate quickly with everyday stress, while others argued that clothing-like interfaces must inherently be designed to withstand wear. Overall, participants regarded context not only as a source of vulnerability but also as a factor shaping expectations of durability.

\subsubsection{Exposure}
\label{section:thematicAnalysisExposure}
Participants (\participants{P01, P06, P08, P11, P17}) considered the interfaces exposure.
They considered resistance to temperature changes (\participants{P11}) and water (\participants{P17}) as factors reducing fragility. 
Meanwhile a lack of water resistance (\participants{P06, P08, P17}), sensitivity to dirt (\participants{P01, P06}) or sand (\participants{P17}), the impact of gravity (\participants{P16}), or the impression the system would break when exposed to the elements in general (\participants{P06}) increased it's perceived fragility.

\subsubsection{Demonstrated Handling}
\label{section:thematicAnalysisHandling}
Participants inferred the fragility of interfaces through the user interaction being demonstrated (\participants{P02-P06, P08-P18}).
Participants concluded that interfaces handled roughly were likely to not be fragile (\participants{P02-P06, P08, P11-P17}), noting \q{I can see that people are putting their whole weight on it so it doesn't look fragile} (\participants{P02}) and \q{it seems sturdy in the way people interacted with it} (\participants{P15}).
This even extended to changing their initial perception with P11 noting \q{I thought maybe pinching it would damage it and then they show us an example of them actually pinching it} and P14 stating \q{I would have thought, oh, this one looks very tear-able, but now he's pinching it and it looks fine}. Demonstrated Handling heavily swayed fragility perception, increasing it when handling was \q{very gentle} (\participants{P05, P16}) or \q{careful} (\participants{P10, P13}), as P10 noted: \q{handling the object seems very careful} which \q{speaks volumes of its fragility}. Contrasting robustness from rough handling, P16 and P17 felt it could increase fragility, with P16 stating \q{the more violent it is, it doesn't necessarily positively contribute to the perception of the things being solid} and P17 noting \q{I'm afraid that some of these will maybe break, especially if people just put all their weight on it}. Participants relied on demonstrated handlings for judgment, with P12 stating \q{it depends maybe on the example that's shown in the video} and P04 noting \q{unless there's more of the video, are they going to show them like batting it with a sledgehammer or something?}. Some expressed doubt about permissible interactions and risk of breakage when not shown (\participants{P11, P13, P15, P16, P18}).

\subsubsection{\finalChanged{Similarity to Existing Objects}}
\label{section:thematicAnalysisSimilarity}
Participants extrapolated interface fragility from their resemblance to familiar systems (\participants{P02-P18}). Some (\participants{P08-P10, P12, P14}) noted that an interface looked like a \q{toy} and was thus likely sturdy. Others compared interfaces to a handfan (\participants{P10:} \q{It's fragile because it's a fan, and fans like this, they're easily broken.}, \participants{P16, P17}), Lego (\participants{P04, P05, P10, P13, P17:} \q{Well, the Lego blocks are notoriously very, very, very solid.}), accordions (\participants{P12, P15}), and keyboards (\participants{P05, P13, P15}). Touchscreens and smartphones were the most referenced (\participants{P02-P04, P06-P07, P09-P14, P18}), with participants noting interfaces could \q{have the same vulnerability of a phone screen if you were to drop your phone} (\participants{P11}), highlighting the role of perceived similarity to existing technology. P15 said that interface P resembled a finger, suggesting this uncanniness may promote careful handling: \q{you wouldn't try and break someone's finger}

\subsubsection{Interface Creation}
\label{section:thematicAnalysisCreation}
\finalChanged{Participants saw interfaces as less fragile, when trusting the designers’ intentions and creation process (\participants{P02, P08, P11, P13, P15, P17}).}
They noted that certain interfaces were likely to have been \q{tested} (\participants{P02}) and had taken considerable time to be created (\participants{P08}).
Participants highlighted some interfaces were \emph{made to be} \q{moved around, thrown around} (\participants{P08}), \q{withstand certain situations} (\participants{P11}) and \q{handled a bit roughly} (\participants{P17}).
They argued an interface couldn't be very breakable because \q{otherwise they wouldn't be designed that way} (\participants{P13}) and stated \q{I kind of trust pretty hard that it's been designed in a way that I cannot reach limits of breaking} (\participants{P15}). Participants also highlighted that some designs may invite improper use (\participants{P11}) and interfaces \q{may break because [they are] not made for that} (\participants{P17}).
Similarly, participants were concerned about the designed interaction limits being unclear, noting \q{that's not what you're intended to do, but might be tempted to do} (\participants{P13}) and \q{I wouldn't know how hard I'm allowed to pinch or raise it} (\participants{P15}).

\subsection{User Factors (Users’ internal processes or interactions with the interface)}

\begin{tcolorbox}\textbf{Summary:}
Participants’ perceptions of fragility were influenced by internal processes and interactions. Feelings toward a device affected judgments, with enjoyment or trust reducing perceived fragility or the opposite (6). Understanding an interface’s functions boosted confidence, while unclear interactions raised concerns (7). Safety considerations were noted by seven participants (7). Interaction possibilities (10), including pinching, twisting, pushing, pulling, picking up, or avoiding parts, shaped judgments, as did the potential for accidental, intentional, or improper damage, including by children or pets. Overall, fragility perception depended on users’ feeling, understanding, interaction strategies, and self-assessed likelihood of damage.
\end{tcolorbox}

\subsubsection{Feelings}
\label{section:thematicAnalysisFeelings}
Participants noted that \finalAdd{their} feelings toward a device influenced perceived fragility (\participants{P01, P08, P10, P12, P14, P18}). P01 stated, \q{I do feel like there are some biases, mind you, whether I like the thing or if I don't}. Liked interfaces were often seen as less fragile, described as \q{resourceful} (\participants{P14}), while disliked ones were judged more fragile, called scary (\participants{P08, P18}) or frustrating (\participants{P12, P18}), with P01 noting, \q{the fact that I don't like it makes me want to put it into the more fragile}. Inexplicable distrust also influenced perception. P01 noted an interface had \q{that thing to it}, P08 said \q{I'm going with my gut. Honestly, I don't very much trust this} and P12 remarked \q{something's not working like I would like to}. This shows that fragility perception is not entirely rational \finalAdd{in some cases. However participants articulated clearly when this was the case and combined their abstract feelings with concrete factors to determine fragility.}

\subsubsection{System Understanding}
\label{section:thematicAnalysisUnderstanding}
Participants commonly tried to deduce the systems functionalities and purpose.
They noted, when the permissible interactions were unclear (\participants{P06, P07, P09, P13, P15, P17, P18}) and highlighted that this would lead to increased damage potential with P15 noting \q{I really did not understand how it works or how it moves or how I might break it}.
Similarly, an increased understanding or comprehension of how an interface worked technically increased the participants confidence \finalChanged{and reduced their fragility perception (\participants{P04, P06-P13, P15, P17}).}
\finalChanged{This suggests, that interfaces that make their mechanism very clear, increase the users confidence in interaction with them, and reduce perceived fragility.}

\subsubsection{Potential Injury}
\label{section:thematicAnalysisInjury}
Participants made explicit reference to the potential of harm (\participants{P07, P08, P10-P12, P15, P18}). Assessing the interface as not fragile, participants (\participants{P10, P12, P15}) noted it was more likely they'd get harmed than the interface stating \q{I'm more worried about myself, like I feel fragile \emph{to} the interface} (\participants{P12}) and emphasising \q{it doesn't look fragile. It feels like I will hurt myself before [breaking the interface]} (\participants{P15}). Meanwhile P07 contemplated that an interface felt \q{a bit more dangerous. Like you could actually hurt someone with that and break it} stressing that in some interfaces the results of fragility could be more detrimental than others. P18 considered one interface could \q{cause harm to humans but it could probably scratch the screen [as well]} highlighting that if an interaction interface is hazardous, potential harm is not limited to users.

\subsubsection{Interactions}
\label{section:thematicAnalysisInteractions}
Participants commonly considered which interactions they may be able to perform with an interface when trying to assess it's fragility. 
They noted the ability to pinch (\participants{P02, P04, P08, P10, P11, P13-P15}), twist (\participants{P13, P15, P16}), or push and pull (\participants{P02, P13, P15}) the interface, as well as whether interaction may be limited to a specific mode (\participants{P01, P07, P13, P15-P17}). 
Further, they considered the way the interface could be picked up (\participants{P02, P11, P13, P16}), or to the contrary, if there were parts that were not supposed to be touched (\participants{P09, P12, P13, P15}), as well as whether you use the device directly with your hands (\participants{P02, P16}).

Through imagining possible interactions, participants considered the potential of breaking the interface. They thought about whether they could break the interface accidentally (\participants{P01, P02, P04, P05, P07, P08, P10-P14, P16-P18}), intentionally (\participants{P01, P07, P08, P11, P12, P17}),
through wrong use (\participants{P01, P03, P06, P11, P13, P15, P17}), or at all (\participants{P01, P02, P04, P05, P07, P08, P10-P13, P15, P17, P18}). Participants also identify several ways the interface may end up broken, such as the interface being pierced (\participants{P10, P15}), peeled (\participants{P10}), burst (\participants{P10}), folded (\participants{P07}), deformed (\participants{P15}), ripped (\participants{P01, P03, P08, P13}), torn (\participants{P02, P04, P06-P08, P11, P12, P14, P16, P18}), bent (\participants{P01, P02, P05, P07, P11, P13, P14, P16}), damaged by dropping (\participants{P02, P04, P05, P07, P08, P10, P11, P13, P14, P18}), by pushing (\participants{P02, P07, P15-P17}), by pulling (\participants{P02, P07, P10, P11, P13-P17}), or getting caught on something (\participants{P01, P03, P07, P11}). These considerations, while being influenced by the characteristics discussed in \autoref{section:thematicAnalysisSystem} \emph{System Factors}, also heavily depended on the participants' self-evaluation and their own likelihood of damaging an interface.
Notably, they extrapolated their own damage potential by considering that interfaces may even be broken by children (\participants{P01, P02, P09, P12, P14-P18}) or pets (\participants{P02, P08, P11, P12}).

\section{Study 2 - Impact of Fragility Factors on Interaction}
\label{section:study2}

To investigate some of the factors identified in the first study and explore %
their effect on physical interaction, 
we conducted a study where participants interacted with SCIs.

\subsection{Selection of Fragility Factors}

\finalAdd{Study 1 revealed a number of factors that appear to shape how people perceive fragility. Investigating this entire landscape in a controlled experiment would require exploring a very large combinatorial space. Such a space cannot be covered within a laboratory protocol because each additional factor multiplies the number of conditions and extends the duration of the study. This would exceed feasible sample size, increase participant fatigue, and compromise data quality. Therefore we selected four to examine closer due to their prevalence in study 1 (see \autoref{table:factorMentions}) and importance to SCI research. }

\begin{itemize}
    \item \finalChanged{\emph{Material} was universally considered by participants as a main factor in the perception of fragility.}

    \item \finalChanged{\emph{Composition} was universally considered by participants. We focus particularly on the amount of parts and consequently their connection as they were highlighted repeatedly as crucial elements of composition.}
    
    \item \emph{Movement} was universally considered by participants and presents a main component of SCIs. Movement and deformation caused significant perception changes. We therefore examine both a shape-changing movement and independent movement that does not include shape-change.
    
    \item \emph{Demonstrated Handling} \finalChanged{was considered by the majority of participants. Crucially, participants revised their initial perceptions due to the handling of objects shown to them.}

\end{itemize}

\subsection{Research Hypothesis and Question}

We hypothesize that the selected factors will shape users’ perceptions of fragility in shape-changing interfaces: \textit{\finalChanged{H:} Material, Composition, Movement and Demonstrated Handling each significantly influence users’ perceptions of fragility in SCIs.} In addition to testing this hypothesis, we sought to understand how fragility perceptions and these factors affect interaction. We therefore posed the following research question: \textit{\finalChanged{RQ:} In what ways do perceived fragility and the implemented factors (Material, Composition, Movement and Demonstrated Handling) influence users’ interaction strategies with SCIs?} We thus analysed participants’ verbal reasoning and movements captured in video recordings to examine how the factors shaped the way they handled the prototypes.

\begin{figure*}[t]
    \centering
    \includegraphics[width=\textwidth]{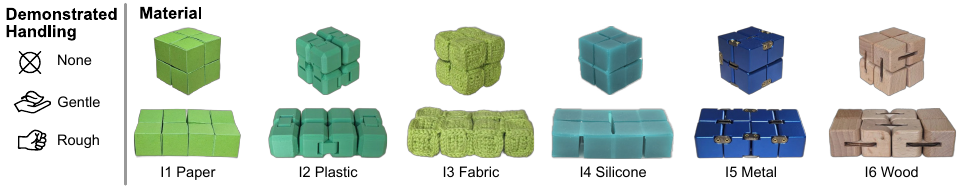}\\
    \caption{Infinity cube implementations used in the experiment including their material, depicting them in a folded and flat state. The 3 levels of demonstrated handling when presenting the cube to participants are summarised on the left.}
    \Description{Infinity cube implementations used in the experiment including their material, depicting them in a folded and flat state. The 3 levels of demonstrated handling when presenting the cube to participants are summarised on the left.}
    \label{figure:infinityCubeTable}
\end{figure*}

\subsection{Implementation of Fragility Factors}
To implement varying levels of these factors, we created two types of objects: Infinity Cubes and Folding Cubes.

\label{section:ICintroduction}
\emph{Infinity Cubes (ICs)} are an established mechanical puzzle toy consisting of 8 interconnected cubes that can be turned over infinitely %
(\autoref{figure:infinityCubeTable}).
We utilise them to test the difference of materials as well as 
demonstrated handling. The materials selected were discussed by participants (\autoref{section:thematicAnalysisMaterial}) - paper, plastic, fabric, silicone and metal (Figure~\ref{figure:infinityCubeTable}).
\finalChanged{We also include wood as a commonly used material, since its absence in our previous results is likely due to  none of the shown interfaces incorporating wood. We exclude glass due to constraints in producing ICs in glass.
The metal and wood IC were purchased %
while we fabricated the others.}
All ICs have a side length of 4-5 cm when folded.
While the hinges connecting the individual cubes inevitably vary due to material constraints (e.g. metal and plastic do not bend, therefore hinges need to be included), this is indicative of the material itself. To examine the impact of demonstrated handling, we test three levels between participants:

\begin{itemize}
    \item None: The IC is not touched by the experimenter but revealed from under a cover.
    \item Gentle demonstration: The experimenter %
    cautiously places the IC in front of the participant.
    \item Rough demonstration: The experimenter roughly places the IC in front of the participant.
\end{itemize}

\begin{figure*}[h]
    \centering
    \includegraphics[width=\textwidth]{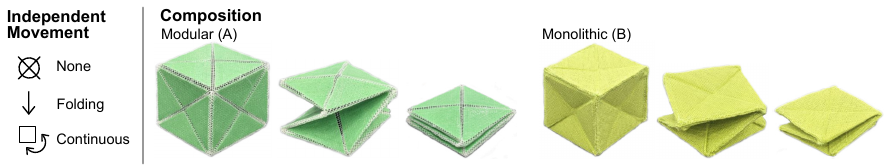}
    \caption{\finalChanged{Folding cube implementations used in the experiment depicting them unfolded, partially folded and fully folded. 
    The 3 levels of independent movement are summarised on the left.
    The 6 implementations are (A) F1: Modular - None, F2: Modular - Folding, F3: Modular - Continuous, (B) F4: Monolithic - None, F5: Monolithic - Folding, F6: Monolithic - Continuous.}}
    \vspace{-1em}
    \Description{Folding cube implementations used in the experiment depicting them unfolded, partially folded and fully folded. The 6 implementations are F1: Modular - None, F2: Modular - Folding, F3: Modular - Continuous, F4: Monolithic - None, F5: Monolithic - Folding, F6: Monolithic - Continuous.}
    \label{figure:foldingCubeTable}
\end{figure*}

\emph{Folding Cubes (FCs)} are origami-inspired, with four triangles per side and foldable flat in any direction. To test the effect of composition and movement we created 6 iterations of folding cubes implementing 2 levels of composition and 3 levels of movement (\autoref{figure:foldingCubeTable}). 
\finalChanged{Composition compares a modular structure presented through a cube showing the individual triangles making up the cube connected through a netting fabric (A) and a monolithic structure, presented through a cube covered in opaque light green fabric (B).} The movement levels presented are:
\begin{itemize}

    \item None: The FC does not move autonomously.
    \item Folding: The FC folds downwards once after being revealed from under a cover.
    \item Continuous: The FC moves around the table in a circular pattern.
\end{itemize}

The folding movement was achieved through strings attached to 2 sides of the cube that were manually pulled downwards through a pedal operated by the experimenter. %
To maintain consistency, the \emph{None} condition featured the same strings. 
While the \emph{Continuous} condition cube similarly had strings affixed to its side, they were not connected to the table to allow the cubes' continuous motion.
The continuous movement was achieved through a magnet sewn into the centre of the cubes bottom side, which attached to a magnet affixed to a motor in the box the objects were presented on.
The motor  
\finalChanged{moved the cube back and forth on a circular path.}

\subsection{Experimental Design and Procedure}
\label{section:study2experimentalDesign}

After giving informed consent, participants completed a demographic survey on a laptop, including age, origin, occupation, handedness, and hand span. They rated 3 questions about their experience with technology on 7-point Likert scales (same as study 1, see \autoref{figure:demographicLikert}, top 3). Participants removed hand jewellery and put on POV camera glasses before stepping into the study space.

\finalAdd{While \emph{Material}, \emph{Movement} and \emph{Composition} were tested within-subjects, \emph{Handling} was tested between-subjects to clearly separate its effect from material differences between the ICs.}
The 6 ICs and 6 FCs were presented in an alternating sequence (IC-FC-IC-FC…), with a balanced Latin square applied within each type to control for order and carry-over effects. Each object appeared once in each of the 12 serial positions across participants within each ``demonstrated handling'' condition.
To allow consistent presentation of the FCs, they were all presented by lifting a cover.
Participants were instructed to demonstrate all possible interactions they could think of. Between objects, participants were asked to step out of the recording space and work on a digital jigsaw puzzle. 

After interacting with the 12 objects, participants %
rated the objects on individual 7-point Likert scales from `1 - Not at all fragile' to `7 - Very fragile' 
and ranked the 6 ICs and 6 FCs from least to most fragile. 
They were then asked to explain their reasoning.
Finally, they were asked to rate the question \q{How would you rate your likelihood of accidentally causing damage (e.g. dropping, breaking or misusing) to something?} on a 7-point Likert scale.

\begin{figure}[h]
    \centering
    \includegraphics[width=\linewidth]{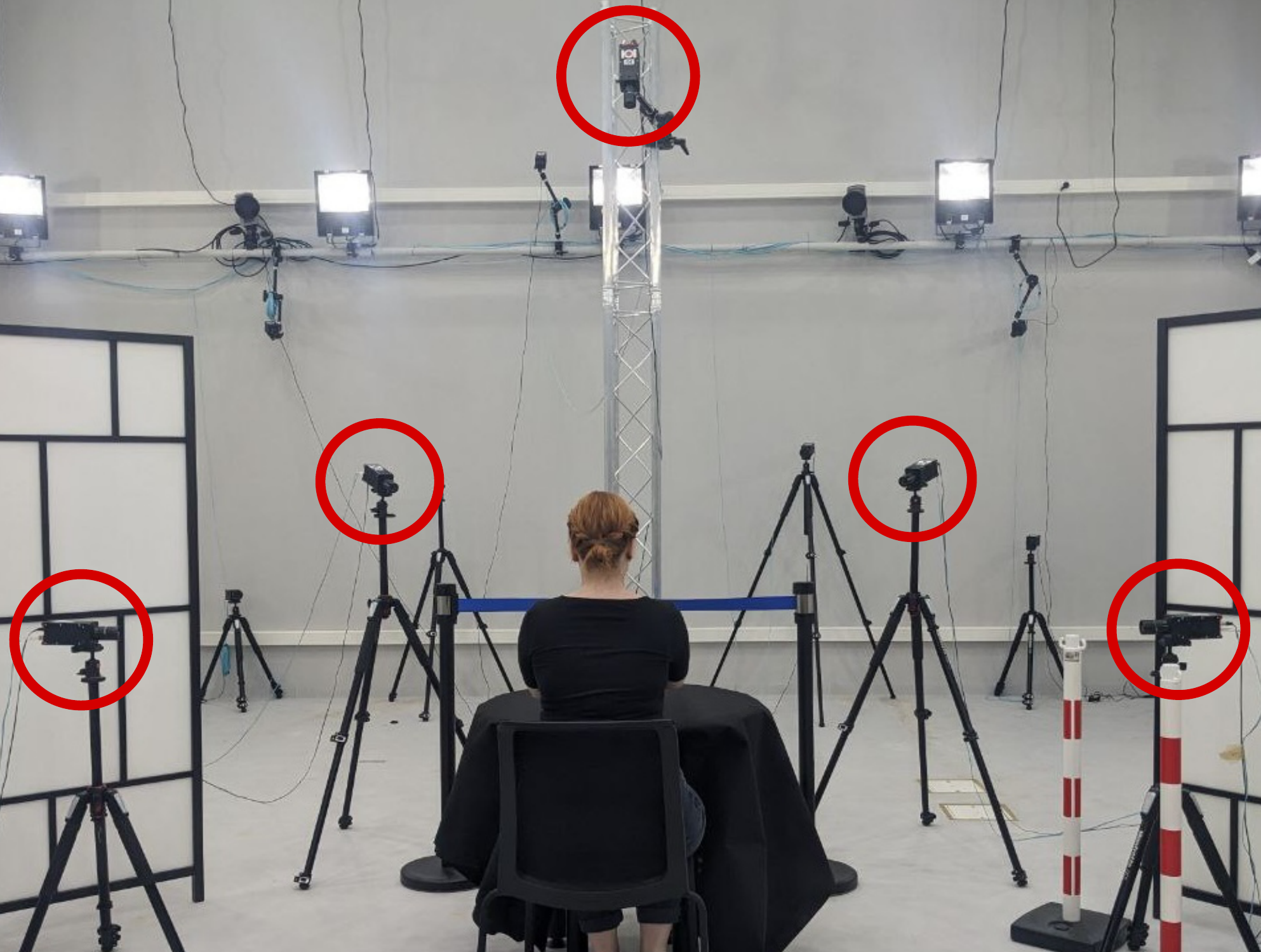}
    \caption{Participant seated in front of a table in the kinovis \cite{kinovis2023} platform, 5 cameras (highlighted in red: 4 pointed towards the table, 1 top-down) and 1 POV camera (worn as glasses). The experimenter (not shown) was stood across the table from the participant.}
    \Description{Set up of the 2nd study: Participant seated in front of table, 5 cameras (highlighted with red circles: 4 pointed towards the table, 1 top-down) and 1 POV camera (worn as glasses). The experimenter (not shown) was stood across the table from the participant.}
    \label{fig:xpsetup}
\end{figure}

\subsection{Capture apparatus}
We used 5 calibrated %
cameras %
(Figure~\ref{fig:xpsetup}) 
placed 1.5 m away from the manipulation table centre.
Two cameras were positioned on either sides of the scene at table level looking horizontally, 
two at 45 degrees on either sides of the subject looking down from 60 cm above the table 
and one was positioned above the table looking down nearly vertically. 
The capture volume was roughly a \finalChanged{50 cm cube}.
We also captured participants' viewpoint with Pupil Labs Invisible glasses.

\subsection{Participants}

We recruited 36 participants via email. 
Age was reported in predefined ranges: 18–24 years (9), 25–34 years (14), 35–44 years (7), 45–54 years (3), and 55-64 years old (3).
21 participants self identified as women and 15 as men.
33 were right-handed and 3 were left-handed.
Their hand-span ranged from 17 cm to 24 cm (\statinfo{M=20.57, SD=1.78}).
Occupations varied from Students (13), Technology or IT (12) and Government or Public Service (10) to Retail or Customer Service (1). Participants self reported a rather high skill level with technology (\statinfo{M=5.22, SD=1.05}), technology usage frequency (\statinfo{M=6.11, SD=0.92}), confidence in learning new technology (\statinfo{M=5.53, SD=1.03}) and a rather low likelihood to accidentally cause damage to something (\statinfo{M=3.94, SD=1.67}). Participation was compensated with a 15 \EUR{} voucher.

\section{Study 2 - Results}

\subsection{Perceived Fragility}

\paragraph{Fragility Ratings and Rankings}
\finalDel{To examine the general perception of fragility as well as the fragility of the presented objects in relation to one another, participants were asked to rate the objects' fragility on a 7-point scale (see Figure 11) and rank all ICs and FCs among themselves from least (1) to most (6) fragile (see Figure 12).}

\begin{figure*}[h]
    \centering
    \includegraphics[width=\linewidth]{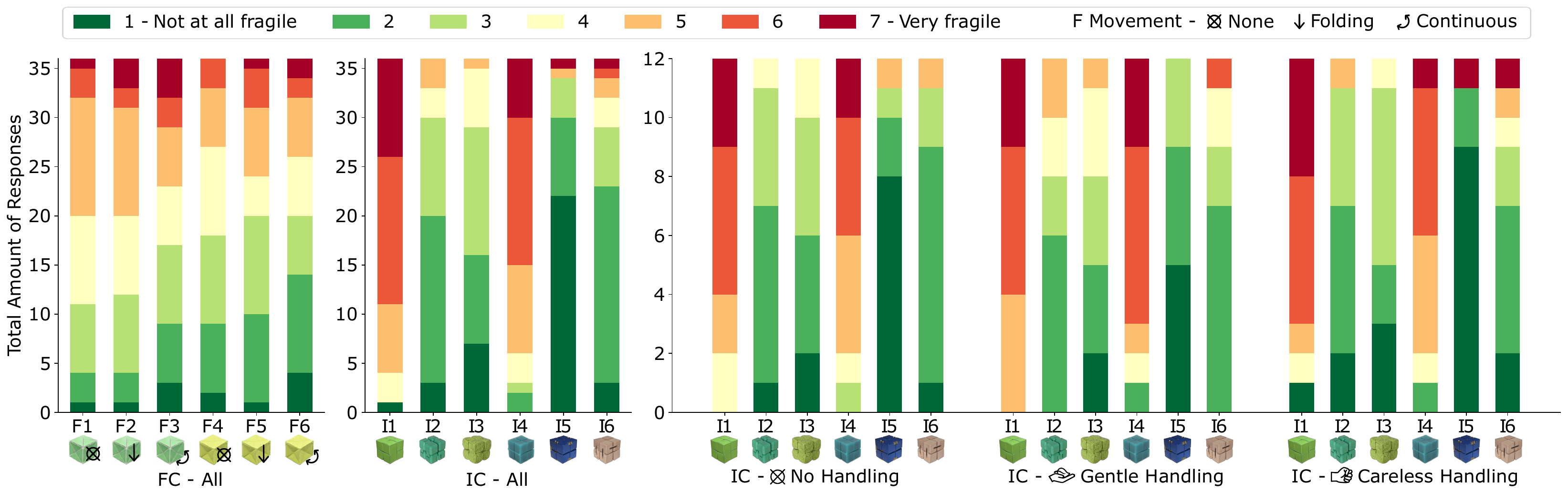}
    \caption{Fragility rankings for the folding cubes (FC) and infinity cube (IC). 
    IC rankings are first presented combining all presentation groups before being separated by how they were handled by the experimenter.}
    \Description{Fragility rankings for the folding cubes (FC) and infinity cube (IC). 
    IC rankings are first presented combining all presentation groups before being separated by how they were handled by the experimenter.}
    \label{fig:likertFragilityFC}
\end{figure*}

\begin{figure}[h]
    \centering
    \vspace{1em}
    \includegraphics[width=0.8\linewidth]{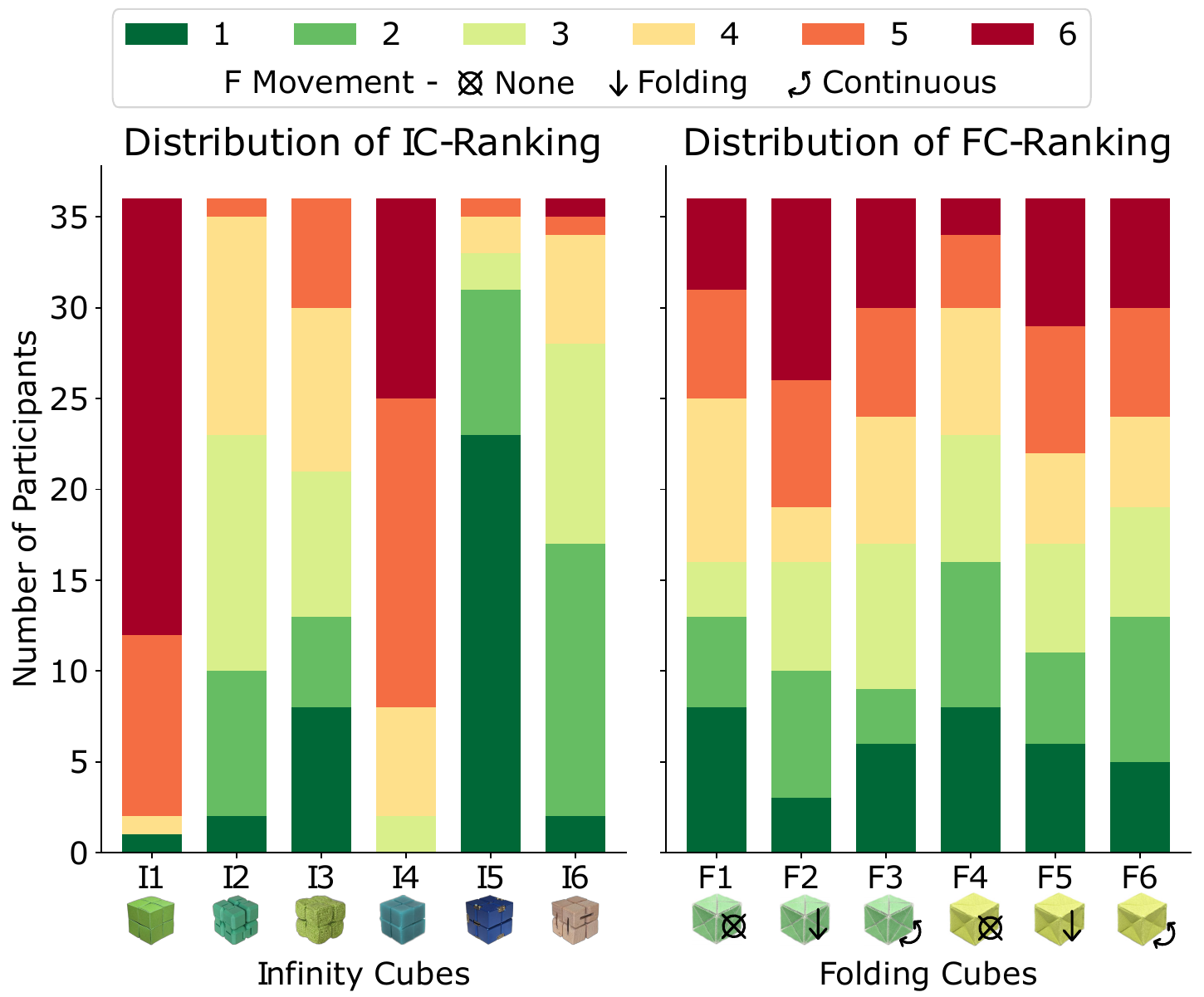}
    \caption{Distribution of IC rankings (left) and FC rankings (right), with objects being rated from least (1) to most (6) fragile within their object type.}
    \Description{Distribution of IC rankings (left) and FC rankings (right), with objects being rated from least (1) to most (6) fragile within their object type.}
    \label{fig:fragilityRanking}
\end{figure}

For the independent fragility \emph{rating} of FCs \finalAdd{(\autoref{fig:likertFragilityFC})} a Friedman test revealed a significant overall difference in fragility ratings across objects ($\chi^2(5) = 15.16, \; p = 0.01$).
However, post-hoc pairwise Wilcoxon comparisons with Bonferroni correction indicated that none of the pairwise differences reached significance after correction. 
Similarly, there was no significant difference between object \emph{rankings} for the FCs  ($\chi^2(5) = 5.94, \; p = 0.31$\finalAdd{, \autoref{fig:fragilityRanking})}.

For the independent fragility \emph{rating} of ICs a Friedman test revealed a significant overall difference in fragility ratings across objects \finalChanged{($\chi^2(5) = 116.04, \; p < 0.01$}, see \autoref{fig:likertFragilityFC} IC -All).
Post-hoc analysis revealed I1 and I4 were rated as significantly more fragile than I2, I3, I5 and I6 while I5 was rated as significantly less fragile than all other ICs (\autoref{appendix:study2posthoc} \autoref{table:ratingICpairwise} shows detailed results).
Equally, there was a significant difference of IC \emph{rankings} \finalChanged{($\chi^2(5) = 112.97, \; p < 0.01$}\finalAdd{, see \autoref{fig:fragilityRanking}, \autoref{appendix:study2posthoc} \autoref{table:rankingICpairwise} shows detailed results)}, with post-hoc analysis revealing the same differences in perceived fragility as the independent ratings. This indicates a significant effect of \emph{Material} on perceived fragility. %
However, fragility ratings between \emph{Demonstrated handling} groups showed no significant difference ($H(2) = 1.61$, $p = 0.45$ (\autoref{fig:likertFragilityFC}). %

\paragraph{Participant Reasoning}

\finalChanged{
For both the ICs and FCs, most participants cited \emph{material} as a key factor, explaining both higher fragility (\participants{I1, P20} \q{it is made of paper. It probably can't withstand a lot of stress}) and lower fragility (\participants{I5, P31} \q{very strong material}). 
Some noted specific aspects like \emph{flexibility} or the 3D-printed nature of I2. 
While many found the thick fabric of F4-F6 (monolithic) \q{stronger in general} (\participants{P7}), others described it as \q{less durable} (\participants{P13}) than the mesh connecting F1-F3 (modular). \\
Participants reflected on their own \emph{interactions} (\participants{I6, P11} \q{it was already difficult to manipulate and I think if I had put much more pressure I might have broken it}) and general interaction possibilities (\participants{I3, P35} \q{you could unravel or cut it pretty easily}). 
For the ICs nine participants noted intentional or accidental breaking (\participants{I1-6, I4-4, P11 broke both}). 
Two participants limited interaction out of fear of breaking the objects: P17 explained they were afraid to \q{destroy some of them by mistake, not knowing whether the destruction of the object can be part of the possible interaction}, and P26 noted \q{I did not want to break them so I didn't really test their fragility}.
Participants also reported being less certain of \q{how to manipulate them [FCs] than the [ICs]} (\participants{P35}), which helps explain volatility in ratings. \\
Composition was considered with a particular focus on \emph{connections}.
Joints between IC cubes, were widely seen as weak points, while the overall cube structure reduced perceived fragility.
Similarly, FC connections, both between triangle surfaces and to the table, influenced judgments: P6 found the mesh in F1-F3 \q{well connected}, while P21 noted F3-F6 were \q{more solid} because \q{the triangles are directly connected to each other}.
Structure was considered to reduce fragility due to the FC monolithic fabric providing a \q{protective outer layer} (\participants{P23}).\\
For the FCs \emph{movement} and deformation were important cues. P23 stated the folding motion made the cube appear robust, whereas others found it \q{very intimidating} (\participants{P13}) or considered the object \q{more fragile cause it folds} (\participants{P36}). P29 noted that a pre-folded cube \q{seemed a little less fragile when I folded it because it easily took shape}. 
The continuous movement of F3 and F6 was seen as increasing fragility (\q{the fact that it moves gives the feeling that we might block its motion and damage it} \participants{P23}). \\
Finally, when discussing perceived fragility, 
participants considered the presence of specific parts 
like the strings and magnets facilitating movement in the FCs,
as well as factors like weight, 
and environmental factors such as water or light. 
}
\finalAdd{For detailed counts please refer to \autoref{appendix:study2reasoning}.}

\subsection{Behaviour Analysis Extracted from the Videos}
Interactions with the objects varied widely, with times ranging from 12.52s to 344.56s (M=84.02), and no significant differences between objects. 
To examine interaction behaviour in detail, we compiled 432 trial videos combining the six camera views of each trial. Three authors collaboratively coded an initial subset of videos to establish consensus. Two authors then independently coded the remaining FC and IC videos. The full coding scheme is provided in the supplementary material.

\subsubsection{Infinity Cube - Material and Demonstrated Handling}
\begin{figure*}[t]
    \centering
    \includegraphics[width=\linewidth]{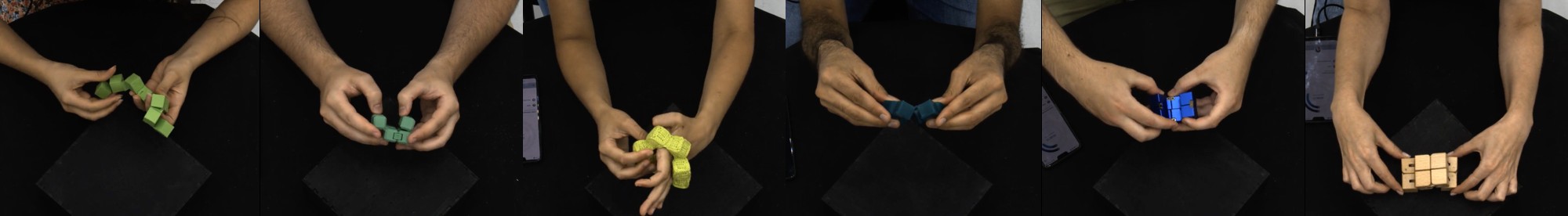}
    \caption{Examples of IC interactions captured by top-down camera. From P1 Paper left to P6 Wood right. 
    }
    \Description{Examples of IC interactions captured by top-down camera. From P1 Paper left to P6 Wood right.}
    \label{fig:ICinteractions}
\end{figure*}

\finalChanged{
In 215/216 of the trials (\emph{36 Participants }$\times$\emph{ 6 Infinity Cubes, 72 trials per Demonstrated Handling}, 
) participants performed the interaction infinity cubes are designed for, of folding the cube into a flat rectangle, folding the rectangle and reforming the object into a cube (which can be performed infinitely giving them their name).
The sole participant who failed to demonstrate this interaction (\participants{P34 None, I4 Silicone}) picked up the cube and threw it into the air, posed the cube on the table and, after noticing the cube could be stretched, tested its limits.
This included a twisting interaction that broke the connection between two cubes, making it impossible for them to then demonstrate the designed interaction.\\
In addition to the designed, and expected, shape change interaction, participants commonly held the cube in 1 hand (including changing the shape with one hand) (None 39/72, Gentle 49/72, Rough 29/72), posed the cube on the table (None 38/72, Gentle 28/72, Rough 20/72), rotated the cube (None 47/72, Gentle 52/72, Rough 60/72), tested the connections of the IC (None 47/72, Gentle 41/72, Rough 53/72) and touched individual cube surfaces (None 15/72, Gentle 26/72, Rough 19/72).
A sample of the most commonly occurring IC manipulations and their frequency across conditions is displayed in \autoref{appendix:study2ICmanipulations}.
}

\paragraph{\finalChanged{Impact of Demonstrated Handling}}
While there are variances in the way participants interacted with the cube depending on \emph{Demonstrated Handling} 
they were not significant with our sample size\finalChanged{, which may be attributed to the limited handling time and variance. This  suggests that}
the impact of demonstrated handling on interactions with an established object like an IC may be more subtle. 

\paragraph{\finalChanged{Impact of Material}}
The material of cubes not only significantly impacted the perceived fragility of the ICs but had a clear impact on demonstrated interactions. Participants broke I1-Paper and I4-Silicone both intentionally (I1 6/36) and accidentally (I1 5/36, I4 4/36).
\finalChanged{This might have impacted perceived fragility, which is supported by P11 explaining they considered I1 fragile since \q{I did destroy it a little even though I was gentle}.}
In contrast P09 noted \q{I destroyed it deliberately. It didn't seem easy to destroy by accident}.
Nonetheless, the perception of fragility did not appear to deter participants from interacting with the objects in potentially destructive ways.
\finalChanged{Further, demonstrated interactions varied between more rigid materials and the flexible fabric and silicone of I3 and I4.
For both, participants were more likely to stretch the cube or squeeze the cube or its parts.
They were the only cubes for which participants demonstrated twisting the cube or attempted to put their hand through its middle (\autoref{fig:ICinteractions} image 3).}

\subsubsection{Folding Cube - Independent movement and Composition}
As shown in Figure~\ref{fig:FCinteractions}, in 133/216 of the trials (62\%) participants performed the folding or unfolding interaction the folding cube was designed for (Figure~\ref{figure:foldingCubeTable}). 

In addition to the designed shape change interaction, 
participants commonly 
grasped the cube (203/216),
rotated the cube (200/216),
interacted with the cube with a single hand (196/216),
or
pressed on the sides of the cube without necessarily folding it (139/216). 
Other interaction varied from 
dropping the cube, 
flicking the cube,
holding it gently, 
interacting with its strings,
lifting or folding its corners, 
shaking it, 
stretching it, 
tapping gently on its sides, 
throwing it, or
moving it on the table. 
As for ICs, participants demonstrated unique interactions like balancing the cube on a body part (head, hand or finger), 
making the cube clap, or 
stroking the cube. 
Fifteen of the most commonly occurring interactions, out of a total of thirty coded interactions, are shown in \autoref{appendix:study2FCmanipulations}. The complete set of coded interactions is provided in the supplementary material.

\begin{figure}[h!]
    \centering
    \includegraphics[width=\linewidth]{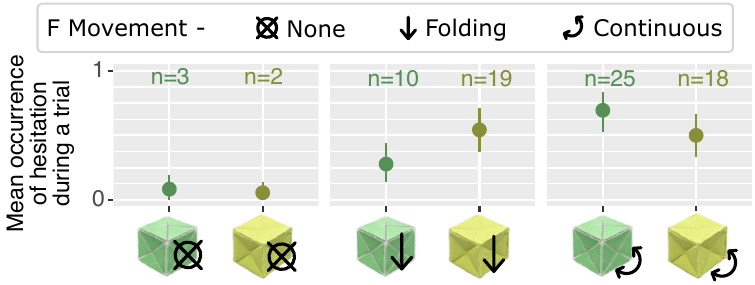}
    \caption{Mean occurrence of hesitations with the FC during a trial.
    0 means participant did not hesitate in this trial, 
    1 means participants hesitated at least once. 
    Above each point representing the mean and the bootstrapped 95\% CI, 
    we show the number of trial where this code appears at least once.}
    \Description{Mean occurrence of hesitations with the FC during a trial.
    0 means participant did not hesitate in this trial, 
    1 means participants hesitated at least once. 
    Next to each point representing the mean and the bootstrapped 95\% CI, 
    we show the number of trial where this code appears at least once.}
    \label{fig:hesitations}
\end{figure}

\begin{figure*}[t]
    \centering
    \includegraphics[width=\linewidth]{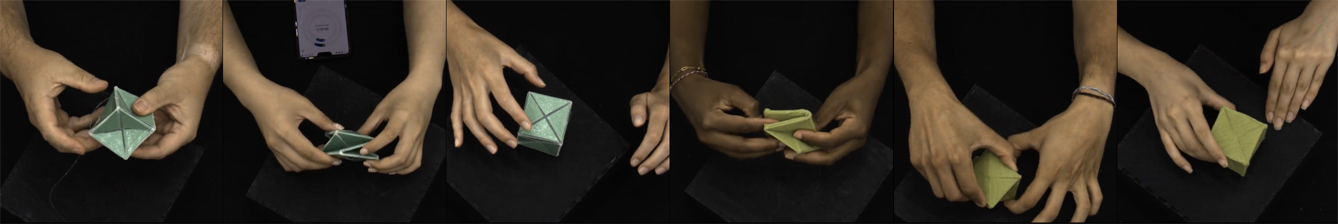}
    \caption{Examples of FC interactions captured by top-down camera. From F1 Modular - None left to F6 Monolithic - Continuous Movement.}
    \Description{Examples of FC interactions captured by top-down camera. From F1 Modular - None left to F6 Monolithic - Continuous Movement.}
    \label{fig:FCinteractions}
\end{figure*}

\paragraph{Impact of Independent Movement}
Independent movement had an impact on participants' hesitations during the trials. 
Figure~\ref{fig:hesitations} shows that they hesitate more during the moving trials (43/72), compared with the folding (29/72) and non actuated trials (5/72). 
\finalChanged{Further, the FC data shows that \emph{system} demonstration through independent movement had an impact on the interactions participants performed. 
Most notably, they performed manipulations that were demonstrated through system actuation. 
During the \emph{continuous movement} trials, participants were more likely to 
translate the cube along the surface (63/72 trials vs. 10/72--\emph{non actuated} and 4/72--\emph{folding}). 
Participants were also more likely to gently flick (25/72 trials vs. 7/72--\emph{non actuated} and 4/72--\emph{folding}) or shake (22/72 trials vs. 14/72--\emph{non actuated} and 6/72--\emph{folding}) the folding cube in the moving trials with continuous movement.
}
In contrast, in the folding condition participants were more likely to twist the cube to fold or unfold it (52/72 trials vs. 35/72--\emph{non actuated} and 25/72--\emph{moving}). 
Presumably in an attempt to unfold the cube, they also lifted or folded the cubes corners (52/72 trials vs. 16/72--\emph{moving} and 35/72--\emph{non actuated}) and stretched the cube more (37/72 trials  vs. 16/72--\emph{non actuated} and 7/72--\emph{moving}) in the folding condition.

\paragraph{Impact of Composition}

\begin{figure}[H]
    \centering
    \includegraphics[width=\linewidth]{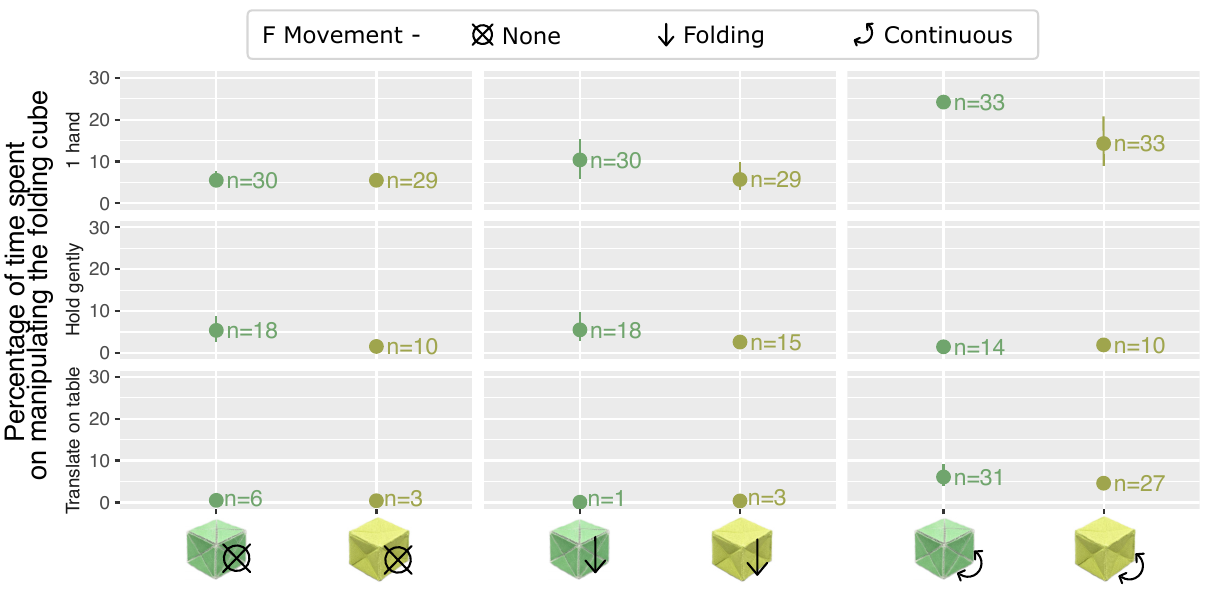}
    \caption{Percentage of time spent on manipulating the folding cube with one hand, holding it gently, or translating it on the table. 
    Next to each point representing the mean and the bootstrapped 95\% CI, 
    we show the number of trial where this code appears at least once.}
    \Description{Percentage of time spent on manipulating the folding cube with one hand, holding it gently, or translating it on the table. 
    Next to each point representing the mean and the bootstrapped 95\% CI, 
    we show the number of trial where this code appears at least once.}
    \label{fig:FCpercentageDurations}
\end{figure}
While there was little difference in interactions between monolithic and modular folding cubes, the mean time (\% of the whole trial) spent on manipulating the cube with a single hand (\autoref{fig:FCpercentageDurations} top) and holding the folding cube gently (\autoref{fig:FCpercentageDurations} middle) shows that participants are more likely to use one hand rather than two hands to manipulate the modular folding cube.
This effect is larger for the continuously moving FC, and neutralised when the FC is not actuated at all. 
As suggested by \citeauthor{10.1145/1753326.1753662}, 
users use fewer hands and less fingers when trying to manipulate a UI more carefully~\cite{10.1145/1753326.1753662}. 
Similarly, participants held the folding cube gently twice as much if it was modular (more than 5\% of the trial) as opposed to monolithic (less than 2.5\% of the trial) in the folding and non actuated conditions. 
This effect stops in the continuous movement condition, 
as participants were more likely to leave the folding cube on the table, 
as shown by the mean time spend on translating the cube on the table (Figure~\ref{fig:FCpercentageDurations} bottom).

\section{Discussion}
\subsection{Framework of Perceived Fragility}
\label{section:fragilityFramework}
\finalChanged{
The results of our studies (summarised in \autoref{table:factorMentions}, visualised in \autoref{figure:thematicAnalysisResults}) establish and reinforce a framework of factors impacting the perception of fragility in SCIs.
In doing so, the framework diverges from previous work on fragility \cite{gorniak2010manipulation} by considering the users subjective perception rather than the physical reality of an interface or object.
Although physical fragility impacts interaction, the users' \emph{perception} similarly guides their action and has previously not been thoroughly considered.
Study 1 examines envisioned perception of fragility in SCIs based on visual cues, while study 2 reaffirms the identified factors apply to practical perceived fragility, highlighting that the same themes
emerge through theoretical examination and physical interaction.
Their combined results highlight that the factors impacting perceived fragility are not only numerous but also diverse, with perceptions being shaped not only by physical properties but also by contextual, emotional, and social factors.
}

\begin{table*}[h]
\centering
  \begin{minipage}{0.5\textwidth}
  \centering
    \subcaption{System factors}
    \begin{tabular}{llrrrrr}\toprule
        \multicolumn{2}{c}{\textbf{Factor}} &\multicolumn{2}{c}{\textbf{Study 1}} &\multicolumn{2}{c}{\textbf{Study 2}} \\\midrule
        \multicolumn{2}{l}{\textbf{Material}} &18 &\cellcolor[HTML]{58a3df}100\% &36 &\cellcolor[HTML]{58a3df}100\% \\
        &Flexibility &13 &\cellcolor[HTML]{90c2ea}72\% &\cellcolor[HTML]{ffffff}20 &\cellcolor[HTML]{a3ccee}56\% \\
        &Consistency/Texture &11 &\cellcolor[HTML]{a6ceee}61\% &\cellcolor[HTML]{ffffff}0 &0\% \\
        &Thickness &10 &\cellcolor[HTML]{b2d5f1}56\% &\cellcolor[HTML]{ffffff}8 &\cellcolor[HTML]{daebf8}22\% \\
        &Specific Materials &18 &\cellcolor[HTML]{58a3df}100\% &36 &\cellcolor[HTML]{58a3df}100\% \\
        \multicolumn{2}{l}{\textbf{Shape}} &17 &\cellcolor[HTML]{64aae2}94\% &\cellcolor[HTML]{ffffff}1 &\cellcolor[HTML]{fbfdff}3\% \\
        &Shape-Change &13 &\cellcolor[HTML]{90c2ea}72\% &\cellcolor[HTML]{ffffff}10 &\cellcolor[HTML]{d1e6f7}28\% \\
        \multicolumn{2}{l}{\textbf{Size}} &15 &\cellcolor[HTML]{7ab6e6}83\% &\cellcolor[HTML]{ffffff}0 &0\% \\
        \multicolumn{2}{l}{\textbf{Weight}} &5 &\cellcolor[HTML]{e9f3fb}28\% &\cellcolor[HTML]{ffffff}9 &\cellcolor[HTML]{d6e8f7}25\% \\
        \multicolumn{2}{l}{\textbf{Composition}} &18 &\cellcolor[HTML]{58a3df}100\% &34 &\cellcolor[HTML]{62a9e1}94\% \\
        &Amount of parts &14 &\cellcolor[HTML]{85bce8}78\% &1 &\cellcolor[HTML]{fbfdff}3\% \\
        &Arrangement &9 &\cellcolor[HTML]{bddbf3}50\% &\cellcolor[HTML]{ffffff}18 &\cellcolor[HTML]{acd1ef}50\% \\
        &Connections &16 &\cellcolor[HTML]{6fb0e4}89\% &28 &\cellcolor[HTML]{7eb8e7}78\% \\
        &Protruding Parts &15 &\cellcolor[HTML]{7ab6e6}83\% &\cellcolor[HTML]{ffffff}0 &0\% \\
        &Filling and Supports &3 &17\% &\cellcolor[HTML]{ffffff}8 &\cellcolor[HTML]{daebf8}22\% \\
        &Specific Parts &17 &\cellcolor[HTML]{64aae2}94\% &\cellcolor[HTML]{ffffff}19 &\cellcolor[HTML]{a7cfef}53\% \\
        \multicolumn{2}{l}{\textbf{Technology}} &10 &\cellcolor[HTML]{b2d5f1}56\% &\cellcolor[HTML]{ffffff}0 &0\% \\
        \multicolumn{2}{l}{\textbf{Cables}} &17 &\cellcolor[HTML]{64aae2}94\% &\cellcolor[HTML]{ffffff}0 &0\% \\
        \multicolumn{2}{l}{\textbf{Function}} &4 &\cellcolor[HTML]{f4f9fd}22\% &\cellcolor[HTML]{ffffff}0 &0\% \\
        \multicolumn{2}{l}{\textbf{Movement}} &18 &\cellcolor[HTML]{58a3df}100\% &16 &\cellcolor[HTML]{b5d7f1}44\% \\
        \multicolumn{2}{l}{\textbf{Quality/Finish}} &12 &\cellcolor[HTML]{9bc8ec}67\% &\cellcolor[HTML]{ffffff}0 &0\% \\
        \multicolumn{2}{l}{\textbf{Longevity}} &12 &\cellcolor[HTML]{9bc8ec}67\% &\cellcolor[HTML]{ffffff}0 &0\% \\
        &Repairability &10 &\cellcolor[HTML]{b2d5f1}56\% &\cellcolor[HTML]{ffffff}0 &0\% \\
        \bottomrule
    \end{tabular}
\end{minipage}%
\begin{minipage}{0.5\textwidth}
  \centering
  \subcaption{Environmental factors}
    \begin{tabular}{llrrrrr}\toprule
        \multicolumn{2}{l}{\textbf{Factor}} &\multicolumn{2}{c}{\textbf{Study 1}} &\multicolumn{2}{c}{\textbf{Study 2}} \\\midrule %
        \multicolumn{2}{l}{\textbf{Usage Context}} &16 &\cellcolor[HTML]{8db976}89\% &\cellcolor[HTML]{ffffff}0 &0\% \\
        &Frequency &6 &\cellcolor[HTML]{d4e5cc}33\% &\cellcolor[HTML]{ffffff}1 &\cellcolor[HTML]{fcfdfb}3\% \\
        &Mobility &10 &\cellcolor[HTML]{b8d4a9}56\% &\cellcolor[HTML]{ffffff}0 &0\% \\
        &Wearables &6 &\cellcolor[HTML]{d4e5cc}33\% &\cellcolor[HTML]{ffffff}0 &0\% \\
        \multicolumn{2}{l}{\textbf{Exposure}} &7 &\cellcolor[HTML]{cde1c3}39\% &\cellcolor[HTML]{ffffff}3 &\cellcolor[HTML]{f5f9f3}8\% \\ 
        \multicolumn{2}{l}{\textbf{Demonstrated Handling}} &16 &\cellcolor[HTML]{8db976}89\% &0 &0\% \\
        \multicolumn{2}{l}{\textbf{Similiarity to Existing Objects}} &17 &\cellcolor[HTML]{86b56d}94\% &\cellcolor[HTML]{ffffff}0 &0\% \\
        \multicolumn{2}{l}{\textbf{Interface Creation}} &6 &\cellcolor[HTML]{d4e5cc}33\% &\cellcolor[HTML]{ffffff}4 &\cellcolor[HTML]{f1f7ee}11\% \\
        \bottomrule
    \end{tabular}
    \vspace{8mm}
    \subcaption{User factors}
    \begin{tabular}{lp{3.7cm}rrrrr}\toprule
        \multicolumn{2}{l}{\textbf{Factor}} &\multicolumn{2}{c}{\textbf{Study 1}} &\multicolumn{2}{c}{\textbf{Study 2}} \\\midrule
        \multicolumn{2}{l}{\textbf{Feelings}} &6 &33\% &\cellcolor[HTML]{ffffff}0 &0\% \\
        \multicolumn{2}{l}{\textbf{Understanding}} &12 &\cellcolor[HTML]{ffe59c}67\% &\cellcolor[HTML]{ffffff}0 &0\% \\
        \multicolumn{2}{l}{\textbf{Potential Injury}} &7 &\cellcolor[HTML]{fffbef}39\% &\cellcolor[HTML]{ffffff}0 &0\% \\
        \multicolumn{2}{l}{\textbf{Interactions}} &18 &\cellcolor[HTML]{ffc938}100\% &\cellcolor[HTML]{ffffff}33 &\cellcolor[HTML]{ffc938}92\% \\
        &Damage Potential &18 &\cellcolor[HTML]{ffc938}100\% &\cellcolor[HTML]{ffffff}33 &\cellcolor[HTML]{ffc938}92\% \\
        \bottomrule
    \end{tabular}
    \label{table:test1}
  \end{minipage}
  \caption{\finalAdd{Tables showing the total count and percentage of participants considering a given factor separated into (a) System factors, (b) Environmental factors and (c) User factors (see also  \autoref{figure:thematicAnalysisResults}) to be influencing their perception of fragility. Study 1 shows factors originating from theoretical considerations envisioning fragility based on visual factors while study 2 shows which factors were considered in a practical interactive setup.}}
  \label{table:factorMentions}
\end{table*}

\finalAdd{
The identified factors mirror those established as core characteristics of SCIs in general (e.g. 
Material \cite{coelho2011shape, sturdee2018classification, qamar2018hcimaterial}, 
Flexibility \cite{roudaut2013morphees, 
10.1145/3173574.3174193,
sturdee2018classification, oosterhout2021deformation, qamar2018hcimaterial}, 
Shape and shape changes \cite{rasmussen2012shapeChangingReview, sturdee2018classification, oosterhout2021deformation,10.1145/3173574.3174193}, 
Size \cite{rasmussen2012shapeChangingReview, sturdee2018classification,oosterhout2021deformation,
10.1145/3173574.3174193}, 
Composition \cite{sturdee2018classification, oosterhout2021deformation},
Technology \cite{sturdee2018classification}, 
Function \cite{rasmussen2012shapeChangingReview}, 
Movement \cite{rasmussen2012shapeChangingReview, roudaut2013morphees, 10.1145/3173574.3174193}, Similarity to Existing Objects (Association) \cite{rasmussen2012shapeChangingReview},  but also factors like  
Interaction \cite{rasmussen2012shapeChangingReview, sturdee2018classification, kwak2014designspace} and 
Feelings \cite{rasmussen2012shapeChangingReview}). 
This alignment both highlights the complexity of fragility perceptions and reaffirms that the identified factors are central to the development of SCIs.
}

\subsubsection{\finalAdd{Envisioned and Practical Perceptions of Fragility}}
\finalAdd{
\autoref{table:factorMentions} shows that there are deviations between factor consideration in the two studies.
While both studies explore the perception of fragility in SCIs, their context significantly impacts the factors considered.
Study 1 presents a theoretical consideration, with participants envisioning fragility based on visual attributes and social cues displayed in a video.
Consequently, usage context and social factors like demonstrated handling are more pronounced. 
Study 2 asked users to interact physically. 
Accordingly, participants primarily considered the interfaces current usage context and their own interactions with the object, suggesting that envisioned perceived fragility provides a more long-term perspective while practical perceived fragility considers an instantaneous contemporary state.
Notably, study 2 is limited by the factors selected while study 1 included a considerably larger variety of interfaces as well as a longer collection of qualitative feedback from participants. 
Nonetheless, the congruence of factors suggests that the visual factors impacting envisioned fragility remain valid for perceived fragility in practical interactions.}

\subsection{\finalChanged{Impact of Fragility Perception and Factors}}
\finalChanged{
Mirroring previous research (\eg \cite{tiab2016understanding, lee2023plantSci}), participant responses in both studies suggest that they were hesitant to interact with interfaces and objects they perceive as fragile (\eg P23 stopping interaction and noting \q{I was afraid to break it}).
This highlights that perceived fragility can \q{discourage usage} \cite{lee2023plantSci} by fundamentally impacting users’ willingness to engage and explore, affecting \textbf{the discoverability of interactions} \cite{mackamul24clarifyingdiscoverability}. 
Given the novelty of SCIs and the lack of clear interaction signifiers, it is essential to support interaction discovery by consciously reducing perceived fragility.
Our findings elucidate that, while the implementation of some factors (like material) yield immediate effects on the perception of fragility, other factors are more dependent on context and the interplay with other characteristics.}

\finalAdd{
However, different characteristics might present conflicts. As  \citeauthor{petersen2020affordances}~\cite{petersen2020affordances} highlight, the addition of covers that obscure actuators and hinges may hide affordances, which aligns with \emph{Understanding} being a factor contributing to reduced fragility.
In contrast, factors like \emph{Technology}, \emph{Cables} and \emph{Composition} increasing perceived fragility suggests that improved clarity through visible parts needs to be balanced with confidence in system robustness produced through covers and monolithic surfaces.}

\subsubsection{\finalChanged{Movement and Composition}}
\finalChanged{Although movement did not significantly impact the subjective perception of fragility (see \autoref{fig:likertFragilityFC}), it did effect actual interaction behaviour in relation to fragility.
Participants were more hesitant to interact with moving interfaces, more likely to mirror movements the device had previously performed independently and deterred from attempting the same deformation if the system was moving in different ways.}
Paradoxically, this reflects the diverging findings of previous work.
Similar to \citeauthor{tiab2016understanding}~\cite{tiab2016understanding} who cite a participants stating \q{It is moving, and if I make it stop I may break it.}, movement increased users' hesitations.
Conversely, independent movements also successfully acted as signifiers for available interactions.
\finalAdd{This impact of movement, including that of different states and their varying fragility, reflects the dynamic nature of shape-changing interfaces and their affordances \cite{follmer2015dynamic}.
Similarly to shape-changes leading to differing perceptions of affordances, they result in differing perceptions of fragility. 
Accordingly, there may be a correlation, with the changed perceived fragility informing the perception of afforded interactions.}
\finalAdd{
It is therefore likely the lack of impact of movement on fragility perceptions in study 2 does not reflect a lack of relevance but rather may be attributed to other factors.
This especially appears to be the case since, while all participants were presented with intact cubes, the modular FC connected through mesh fabric factually was more fragile, with multiple replacements being required throughout the experiment.
Potentially differences in composition were not pronounced enough, given the repetitive nature of the experiment.
Since the study was intentionally designed to not mention fragility before participants had interacted with all interfaces to avoid biasing them, participants rated all interfaces at once.
This alongside the FCs similar visual appearances might have contributed to participants struggling to differentiate between different cubes when subsequently evaluating their fragility, which is anecdotally supported by multiple participants asking for clarifications of which FC was which when filling out their questionnaires.
}

\subsubsection{\finalAdd{Demonstrated Handling}}
\finalAdd{While demonstrated handling was a common and persuasive factor in study 1, this was not replicated in study 2. Given that shape changes that were demonstrated by the system rather than the experimenter did influence behaviour, it is likely that the lack of impact in study 2 can be attributed to the the type of demonstrated handling.
The presentation of objects might have been too short and differences between conditions not severe enough to change behaviour, especially given that no interactions other than placing the object down were demonstrated.
The choice to select short handling was made to replicate both, common experimental conditions and brief rough interactions that appeared to have a considerable effect on perceived fragility in study 1.
We refrained from longer demonstrated handling to reduce bias on participants towards specific interactions, which might have rendered the effect of demonstrated handling moot with out sample size and interfaces.}

\subsubsection{\finalAdd{Material and Colour}}
\finalAdd{
Material has been widely considered in the implementation of SCIs \cite{qamar2018hcimaterial, alexander2018challenges} and emerged as a key factor in the perception of fragility in our studies.
We produced ICs in 6 materials in study 2 to examine their perception and impact.
Given that colour can impact the perception of materials \cite{schmidt2025materialdimensions} we presented \emph{I6}, which was made out of wood, in a natural beige, since there is a clear association between colour and material that is not as evident in the other materials.
Due to limitations in our production capacity and less clear colour associations, the other ICs were produced in different shades of green and blue to increase visibility.
While colour contributes to the identification of materials, it alone is not crucial to material recognition \cite{xiao2025material, zaidi2011visual}.
Materials that commonly occur in various colours are not likely to be perceived wrong due to artificial colours, as evidenced by all our participants being able to clearly articulate the materials they were confronted with.
Further, different colours appears to have helped participants differentiate more clearly between interfaces, as they commonly referred to them by colour in the subsequent evaluations.
Nonetheless, beyond material identification colour also contributes to the perception of object properties in tangible interfaces \cite{loeffler2018multimodal}.
In our study, due to the similarity in other visual factors such as shape and size, the variance in colour is likely not congruent to expected haptic properties, limiting the perceptual effect of colour differences \cite{loeffler2018multimodal}.
Despite these limitations, we cannot be certain that differences in colour did not impact our participants' perception and behaviour.
}

\subsection{Importance of this Work}
Beyond its potential impact on interaction discoverability, the perception of fragility in SCIs has significant implications.
Given that \q{current systems are often not \emph{robust} enough for in-depth evaluations} \cite{alexander2018challenges} it is essential to formalise factors to allow researchers and designers to make informed decisions. We have provided a framework (\finalAdd{\autoref{section:fragilityFramework}}, Figure \ref{figure:thematicAnalysisResults}), enabling researchers and designers to make educated choices to manage fragility perceptions. Our findings also suggest several design directions that researchers and designers can explore to shape perceived fragility in SCIs:

\begin{itemize}

\item \textbf{\finalChanged{Material choice as an important cue:}} Material consistently dominated fragility judgments \finalAdd{in our studies}. Lightweight and flexible materials (e.g., paper or silicone) were perceived as more fragile than rigid materials like wood or metal. Designers can leverage material properties to signal durability or delicacy depending on the intended interaction, encouraging careful exploration or vigorous interaction.

\item \textbf{Structural features and connections as signals of robustness:} Participants frequently identified joints, seams, and strings as potential weak points. Visible connection elements strongly influence perceived fragility. Designers should carefully consider how structural features are presented; exposed or tenuous connections may signal fragility, while reinforced or concealed connections can communicate robustness.

\item \textbf{Ambiguity in dynamic behaviour:} Movement cues produced divergent interpretations. Dynamic behaviour alone cannot reliably convey fragility, durability or interactivity. Designers should either supplement movement with clear signifiers like material cues to reduce ambiguity or exploit it to spark curiosity and varied interactions.

\item \textbf{Fragility does not deter engagement:} Perceived fragility did not prevent participants from engaging with the prototypes. Fragility framed how interaction occurred, fostering attentive, exploratory, and playful engagement. Designers can treat fragility as a resource rather than a constraint.

\item \textbf{Supporting interaction discoverability:}
\finalChanged{While perceived fragility does not deter interaction in general, it can influence how users approach unfamiliar SCIs,  making them hesitant to fully explore available interactions.} Designers should provide clear affordances through signifiers, or demonstrations that indicate safe ways to interact, helping users discover all functionalities without fear of damage. Thoughtful scaffolding of interaction discoverability ensures that fragility becomes an informative feature rather than a barrier.

\end{itemize}

\subsection{Limitations}
While our studies provides a valuable first insight into the factors influencing the perception of fragility in SCIs we cannot claim the framework to be complete, since there might be contributing factors that did not get revealed given our sample of interfaces and participants as well as the limited factors explored through physical interaction.
\finalAdd{
Consequently, while participant feedback in Study 2 revealed factors that align with those identified in study 1, even beyond the ones intentionally implemented, other factors are not examined, leaving their pertinence to practical fragility perception to be determined.}
Further, since interaction context inevitably impacts the perception of SCIs \cite{kinch2014benchchange}
it is possible that the context of our user study and being told to interact negates some hesitation since part of the ``responsibility'' is on the research rather than the participant, therefore not accurately reflecting the impact of initial perception of fragility in real life contexts.
Nonetheless, given the variety of interfaces covered in the studies and the concerted effort to include participant from varying backgrounds, we can be reasonably certain that the framework covers the most commonly occurring factors.

\subsection{\finalChanged{Future Work}}
\finalAdd{While the presented studies provide a structural foundation of factors to consider in the envisioned and practical perception of fragility, the impact of individual factors as well as their interplay needs to be further validated.}
\finalChanged{Additionally, despite the difficulty of generalising explorations of interactions with SCIs, future work should examine how perceptual factors beyond perceived fragility impact user behaviour.}
Conventional and commercially available interfaces increasingly diverge from traditional form factors, making the exploration of their perceived fragility essential to understanding their initial exploration.
Such explorations could be utilised to confirm and expand upon our framework.
Finally, fragility only presents a small part of the overall perceptual elements informing our mental models and subsequent interaction with interfaces.
Other factors such as responsiveness \cite{doherty2015keeping}, discoverability \cite{mackamul24clarifyingdiscoverability} or valence \cite{beale2008role} should similarly be explored and considered in correlation with fragility to establish a more universal understanding of the perception of SCIs and how we might utilise it to promote successful user interactions.

\section{Conclusion}
There are consistent factors that are foundational to the perception of fragility of SCIs.
Through two user studies, addressing the theoretical consideration of fragility based on video stimuli and practical interactions and perceptions respectively, we found that, while weighting varies between individual users, and focal points change depending on the context in which an interface is evaluated, the same factors impact the perception of fragility in both theoretical and practical physical interaction.
We offer a framework that collates these factors and organises them into inherent system characteristics, environmental factors as well as user factors in an effort to support conscious consideration of fragility in the design of SCIs and encourage the development of interfaces that invite user engagement.
While some factors generate immediate changes in perception and behaviour by themselves, others need to be further explored to elucidate the interplay of characteristics and their effect depending on context.

\section{Acknowledgment}

This work was supported by French government funding managed by the National Research Agency ANR-21-CE33-0018 (SecondSkin) project as well as under the Investments for the Future program (PIA) grant ANR-21-ESRE-0030 (CONTINUUM).
We gratefully acknowledge the Grenoble Computer Science Laboratory (LIG-Laboratoire d’Informatique de Grenoble) for providing funding for a postdoctoral fellowship, which enabled the completion of this study.
We would like thank the Fablab MASTIC for their support in fabricating the prototypes.

\bibliographystyle{plainnat}
\bibliography{biblio}

\newpage
\onecolumn
\section{Study 2 Posthoc Analysis - Infinity Cube (IC) Fragility likert rating and ranking}
\label{appendix:study2posthoc}
\centering
\begin{table}[H]
\centering
\begin{minipage}{0.48\textwidth}
\centering
\begin{tabular}{llrr}
\hline
Object1 & Object2 & W & p \\ 
\hline
I1 & I2 & 3.00 & \textbf{<0.001} \\ 
I1 & I3 & 0.00 & \textbf{<0.001} \\ 
I1 & I4 & 209.50 & 1.000 \\ 
I1 & I5 & 1.50 & \textbf{<0.001} \\ 
I1 & I6 & 9.00 & \textbf{<0.001} \\ 
I2 & I3 & 184.00 & 1.000 \\ 
I2 & I4 & 8.00 & \textbf{<0.001} \\ 
I2 & I5 & 56.00 & \textbf{0.005} \\ 
I2 & I6 & 100.00 & 1.000 \\ 
I3 & I4 & 2.50 & \textbf{<0.001} \\ 
I3 & I5 & 88.00 & \textbf{0.037} \\ 
I3 & I6 & 209.50 & 1.000 \\ 
I4 & I5 & 5.50 & \textbf{<0.001} \\ 
I4 & I6 & 9.00 & \textbf{<0.001} \\ 
I5 & I6 & 29.50 & \textbf{0.003} \\ 
\hline
\end{tabular}
\caption{Post-hoc pairwise comparison between fragility ratings of ICs with Bonferroni corrected p-values.}
\label{table:ratingICpairwise}
\end{minipage}\hfill
\begin{minipage}{0.48\textwidth}
\centering
\begin{tabular}{llrr}
\hline
Object1 & Object2 & W & $p$ \\
\hline
I1 & I2 & 23.50 & \textbf{<0.001} \\
I1 & I3 & 13.00 & \textbf{<0.001} \\
I1 & I4 & 203.00 & 0.465 \\
I1 & I5 & 14.50 & \textbf{<0.001} \\
I1 & I6 & 35.50 & \textbf{<0.001} \\
I2 & I3 & 322.50 & 1.000 \\
I2 & I4 & 19.50 & \textbf{<0.001} \\
I2 & I5 & 61.50 & \textbf{<0.001} \\
I2 & I6 & 268.50 & 1.000 \\
I3 & I4 & 37.50 & \textbf{<0.001} \\
I3 & I5 & 125.50 & \textbf{0.015} \\
I3 & I6 & 299.00 & 1.000 \\
I4 & I5 & 6.00 & \textbf{<0.001} \\
I4 & I6 & 29.00 & \textbf{<0.001} \\
I5 & I6 & 78.00 & \textbf{<0.001} \\
\hline
\end{tabular}
\caption{Post-hoc pairwise comparison between fragility rankings of ICs with Bonferroni corrected p-values.}
\label{table:rankingICpairwise}
\end{minipage}
\end{table}

\section{Study 2 Participant Reasoning - Factor Counts}
\label{appendix:study2reasoning}
\begin{table}[H]\centering
\begin{tabular}{llllllll}\toprule
&\makecell{I1 Paper} &\makecell{I2 Plastic} &\makecell{I3 Fabric} &\makecell{I4 Silicone} &\makecell{I5 Metal} &\makecell{I6 Wood} \\\midrule
Material &\cellcolor[HTML]{6aa84f}32 &\cellcolor[HTML]{92c07e}21 &\cellcolor[HTML]{8bbc76}23 &\cellcolor[HTML]{b6d5a9}11 &\cellcolor[HTML]{80b569}26 &\cellcolor[HTML]{84b76d}25 \\
\textbf{ } 3D-printed &0 &\cellcolor[HTML]{c4deba}7 &0 &0 &0 &0 \\
\textbf{ } Flexibility &0 &0 &\cellcolor[HTML]{b6d5a9}11 &\cellcolor[HTML]{c8e0be}6 &\cellcolor[HTML]{d6e8cf}2 &0 \\
Connections &\cellcolor[HTML]{cbe2c2}5 &\cellcolor[HTML]{afd1a0}13 &\cellcolor[HTML]{d6e8cf}2 &\cellcolor[HTML]{a0c88f}17 &\cellcolor[HTML]{95c283}20 &\cellcolor[HTML]{95c283}20 \\
Structure &\cellcolor[HTML]{d2e6cb}3 &\cellcolor[HTML]{d6e8cf}2 &\cellcolor[HTML]{cfe4c7}4 &\cellcolor[HTML]{d9ead3}1 &\cellcolor[HTML]{d2e6cb}4 &0 \\
\textbf{ }  Hollow &\cellcolor[HTML]{d9ead3}1 &\cellcolor[HTML]{d9ead3}1 &0 &0 &0 &0 \\
\textbf{ } Thickness &0 &\cellcolor[HTML]{d9ead3}1 &\cellcolor[HTML]{d9ead3}1 &\cellcolor[HTML]{d6e8cf}2 &0 &\cellcolor[HTML]{d9ead3}1 \\
Construction &\cellcolor[HTML]{d9ead3}1 &\cellcolor[HTML]{d9ead3}1 &\cellcolor[HTML]{d6e8cf}2 &0 &0 &0 \\
Weight &\cellcolor[HTML]{d2e6cb}3 &0 &0 &0 &\cellcolor[HTML]{c0dcb6}8 &0 \\
Interaction &\cellcolor[HTML]{b9d7ad}10 &\cellcolor[HTML]{bdd9b1}9 &\cellcolor[HTML]{a4cb94}15 &\cellcolor[HTML]{9dc68b}18 &\cellcolor[HTML]{cbe2c2}5 &\cellcolor[HTML]{a7cd98}15 \\
Environment &\cellcolor[HTML]{d6e8cf}2 &0 &0 &\cellcolor[HTML]{d9ead3}1 &0 &0 \\
\bottomrule
\label{table:study2fragilityReasoning}
\end{tabular}
\caption{Participant reasoning for fragility ratings of ICs.}
\end{table}

\begin{table}[H]\centering
\centering
\begin{tabular}{llllllll}\toprule
&\multicolumn{2}{l}{Modular} & &\multicolumn{2}{l}{Monolithic} & \\
&\makecell{F1 None} &\makecell{F2 Folding} &\makecell{F3 Moving} &\makecell{F4 None} &\makecell{F5 Folding} &\makecell{F6 Moving} \\\midrule
Interaction &\cellcolor[HTML]{91c07e}14 &\cellcolor[HTML]{70ac56}20 &\cellcolor[HTML]{86b970}16 &\cellcolor[HTML]{b8d7ac}7 &\cellcolor[HTML]{81b66a}17 &\cellcolor[HTML]{a2c991}11 \\
Connection &\cellcolor[HTML]{c3ddb9}5 &\cellcolor[HTML]{c9e1c0}4 &\cellcolor[HTML]{b8d7ac}7 &\cellcolor[HTML]{c9e1c0}4 &\cellcolor[HTML]{cee4c6}3 &\cellcolor[HTML]{c3ddb9}5 \\
\textbf{ }\makecell{Wires/Strings} &\cellcolor[HTML]{9cc68b}12 &\cellcolor[HTML]{97c384}13 &\cellcolor[HTML]{bedab2}6 &\cellcolor[HTML]{cee4c6}3 &\cellcolor[HTML]{bedab2}6 &\cellcolor[HTML]{cee4c6}3 \\
\textbf{ } Magnet &0 &0 &\cellcolor[HTML]{bedab2}6 &0 &0 &\cellcolor[HTML]{add09f}9 \\
Structure &\cellcolor[HTML]{cee4c6}3 &\cellcolor[HTML]{d4e7cd}2 &\cellcolor[HTML]{cee4c6}3 &\cellcolor[HTML]{d4e7cd}2 &\cellcolor[HTML]{d9ead3}1 &\cellcolor[HTML]{c9e1c0}4 \\
Materials &\cellcolor[HTML]{9cc68b}12 &\cellcolor[HTML]{a8cd98}10 &\cellcolor[HTML]{b8d7ac}7 &\cellcolor[HTML]{6aa84f}21 &\cellcolor[HTML]{a2c991}11 &\cellcolor[HTML]{9cc68b}12 \\
\textbf{ } Thickness &0 &0 &0 &\cellcolor[HTML]{d4e7cd}2 &\cellcolor[HTML]{cee4c6}3 &0 \\
\textbf{ } Flexible &\cellcolor[HTML]{c3ddb9}5 &\cellcolor[HTML]{d4e7cd}2 &0 &\cellcolor[HTML]{d4e7cd}2 &\cellcolor[HTML]{cee4c6}3 &0 \\
Shape &\cellcolor[HTML]{d9ead3}1 &\cellcolor[HTML]{d9ead3}1 &\cellcolor[HTML]{d9ead3}1 &\cellcolor[HTML]{d9ead3}1 &\cellcolor[HTML]{d9ead3}1 &\cellcolor[HTML]{d9ead3}1 \\
\makecell{Protective Layer} &\cellcolor[HTML]{d4e7cd}2 &\cellcolor[HTML]{d9ead3}1 &\cellcolor[HTML]{d4e7cd}2 &\cellcolor[HTML]{cee4c6}3 &\cellcolor[HTML]{cee4c6}3 &\cellcolor[HTML]{d4e7cd}2 \\
State &\cellcolor[HTML]{d9ead3}1 &\cellcolor[HTML]{cee4c6}3 &\cellcolor[HTML]{cee4c6}3 &0 &\cellcolor[HTML]{bedab2}6 &\cellcolor[HTML]{d4e7cd}2 \\
Movement &0 &\cellcolor[HTML]{b3d3a5}8 &\cellcolor[HTML]{c3ddb9}5 &\cellcolor[HTML]{d9ead3}1 &\cellcolor[HTML]{c3ddb9}5 &\cellcolor[HTML]{bedab2}6 \\
\bottomrule
\label{tab:study2fragilityReasoningFCs}
\end{tabular}
\caption{Participant reasoning for fragility ratings of FCs.}
\end{table}

\section{Study 2 Behavioural Analysis FC Interactions}
\label{appendix:study2FCmanipulations}
\begin{figure*}[h]
    \centering
    \includegraphics[width=\linewidth]{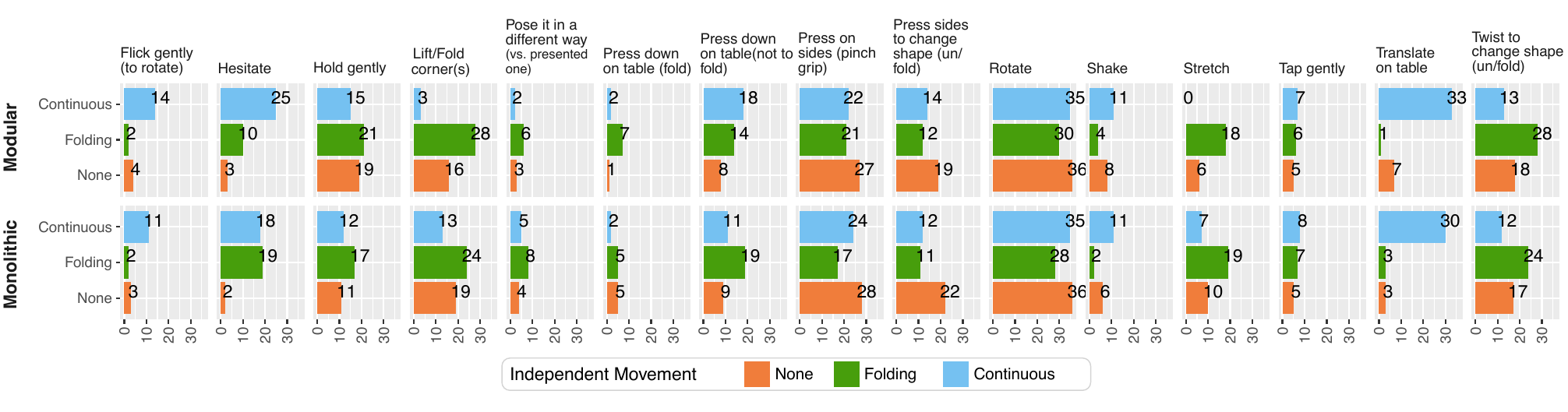}
    \caption{Number of participants that perform a given FC manipulation in the trial at least once separated by \emph{Independent Movement}. %
    }
    \Description{Number of participants that perform a given FC manipulation in the trial at least once separated by Independent Movement.}
    \label{fig:occurences}
\end{figure*}

\clearpage
\section{Study 2 Behavioural Analysis IC Interactions}
\label{appendix:study2ICmanipulations}
\begin{center}
    \includegraphics[width=0.90\linewidth]{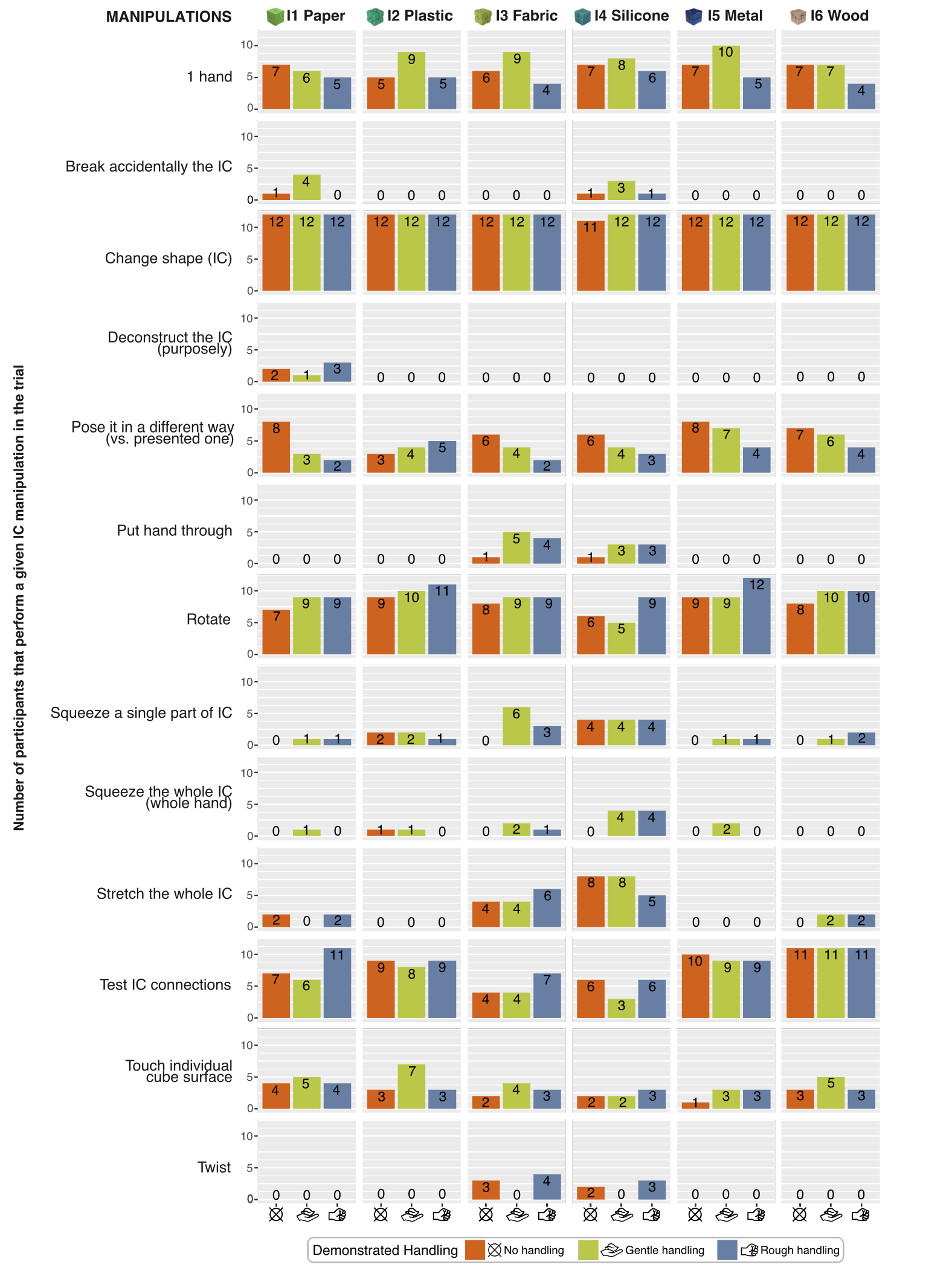}
    \captionof{figure}{Number of participants that perform a given IC manipulation in the trial separated by \emph{Demonstrated Handling}.}
    \Description{Number of participants that perform a given IC manipulation in the trial separated by Demonstrated Handling}
    \label{fig:ICOccurrencesParticipants}
\end{center}

\end{document}